\documentclass[aoas,nameyear,dvips]{arximspdf}
\usepackage{algorithm,algorithmic}
\usepackage{graphicx}
\usepackage{textcomp}

\doi{10.1214/09-AOAS303}
\volume{5}
\issue{2B}
\pubyear{2011}
\firstpage{1229}
\lastpage{1261}

\makeatletter
\newcommand{\eqref}[1]{(\ref{#1})}

\makeatother

\begin{document}
\begin{frontmatter}

\title{A Bayesian approach for inferring neuronal connectivity from
calcium fluorescent\\ imaging data}
\runtitle{Inferring neuronal connectivity from calcium imaging}

\begin{aug}
\author[a]{\fnms{Yuriy} \snm{Mishchencko}\corref{}\ead[label=e1]{yuriy.mishchenko@gmail.com}},
\author[b]{\fnms{Joshua T.} \snm{Vogelstein}\thanksref{t1}\ead[label=e2]{joshuav@jhu.edu}}
\and
\author[a]{\fnms{Liam} \snm{Paninski}\thanksref{t2}\ead[label=e3]{liam@stat.columbia.edu}
\ead[label=u1,url]{http://www.stat.columbia.edu/\textasciitilde liam}}%
\runauthor{Y. Mishchenko, J. T. Vogelstein and L. Paninski}
\affiliation{Columbia University, Johns Hopkins University and Columbia
University}
\thankstext{t1}{Supported by NIDCD Grant DC00109.}
\thankstext{t2}{Supported by a NSF CAREER and McKnight Scholar award.}
\address[a]{Y. Mishchenko\\
L. Paninski\\
Department of Statistics\\
\quad and Center for Theoretical Neuroscience\\
Columbia University \\
1255 Amsterdam Ave \\
New York, New York 10027\\
USA \\
\printead{e1}\\
\phantom{E-mail:} \printead*{e3}\\
\printead{u1}}
\address[b]{J. Vogelstein\\
Johns Hopkins University\\
3400 N. Charles St.\\
Baltimore, Maryland 21205 \\
USA \\
\printead{e2}}
\end{aug}

\received{\smonth{9} \syear{2009}}
\revised{\smonth{10} \syear{2009}}

%
\begin{abstract}
Deducing the structure of neural circuits is one of the central
problems of modern neuroscience. Recently-introduced calcium
fluorescent imaging methods permit experimentalists to observe network
activity in large populations of neurons, but these techniques provide
only indirect observations of neural spike trains, with limited time
resolution and signal quality. In this work we present a Bayesian
approach for inferring neural circuitry given this type of imaging
data. We model the network activity in terms of a collection of
coupled hidden Markov chains, with each chain corresponding to a~%
single neuron in the network and the coupling between the chains
reflecting the network's connectivity matrix. We derive a Monte Carlo
Expectation--Maximization algorithm for fitting the model parameters;
to obtain the sufficient statistics in a computationally-efficient
manner, we introduce a specialized blockwise-Gibbs algorithm for
sampling from the joint activity of all observed neurons given the
observed fluorescence data. We perform large-scale simulations of
randomly connected neuronal networks with biophysically realistic
parameters and find that the proposed methods can accurately infer the
connectivity in these networks given reasonable experimental and
computational constraints. In addition, the estimation accuracy may
be improved significantly by incorporating prior knowledge about the
sparseness of connectivity in the network, via standard L$_1$
penalization methods.
\end{abstract}

%

\end{frontmatter}
%

\section{Introduction}
\label{intro}
Since Ramon y Cajal discovered that the brain is a~rich and dense
network of neurons [Ramon y Cajal (\citeyear{RamonyCajal04,RamonyCajal23})], neuroscientists
have been intensely curious about the details of these networks, which
are believed to be the biological substrate for memory, cognition and
perception. While we have learned a great deal in the last century
about ``macro-circuits'' (the connectivity between coarsely-defined
brain areas), a number of key questions remain open about
``micro-circuit'' structure, that is, the connectivity within populations
of neurons at a fine-grained cellular level. Two complementary
strategies for investigating micro-circuits have been pursued
extensively. \textit{Anatomical} approaches to inferring circuitry do
not rely on observing neural activity; some recent exciting examples
include array tomography [\citet{MichevaSmith07}], genetic ``brainbow''
approaches [\citet{Brainbow07}], and serial electron microscopy
[\citet{Briggman2006}]. Our work, on the other hand, takes a
\textit{functional} approach: our aim is to infer micro-circuits by
observing the simultaneous activity of a population of neurons,
without making direct use of fine-grained anatomical
measurements.\looseness=-1

Experimental tools that enable simultaneous observations of the
activity of many neurons are now widely available. While arrays of
extracellular electrodes have been exploited for this purpose
[\citet{HATS98}; \citet{HARR03}; \citet{Stein04}; \citet{Santhanam06}; \citet{Harris07}], the arrays most
often used {in vivo} are inadequate for inferring monosynaptic
connectivity in large populations of neurons, as the inter-electrode
spacing is typically too large to record from closely neighboring
neurons;\setcounter{footnote}{2}\footnote{It is worth noting, however, that multielectrode
arrays which have been recently developed for use in the retina
[\citet{Berry2004}; \citet{Litke2004}; \citet{Petrusca07}; \citet{PILL07}] or in cell culture
[\citet{Shepard08}] are capable of much denser sampling.} importantly,
neighboring neurons are more likely connected to one another than
distant neurons [\citet{Abeles91}; \citet{Braitenberg1998}]. Alternately,
calcium-sensitive fluorescent indicators allow us to observe the
spiking activity of on the order of $10^3$ neighboring neurons
[\citet{Tsien89}; \citet{ImagingManual}; \citet{CAR03}; \citet{OHKI05}] within a micro-circuit.
Some organic dyes achieve sufficiently high signal-to-noise ratios
(SNR) that individual action potentials (spikes) may be resolved
[\citet{ImagingManual}], and bulk-loading techniques enable
experimentalists to simultaneously fill populations of neurons with
such dyes [\citet{StosiekKonnerth03}]. In addition, genetically encoded
calcium indicators are under rapid development in a number of groups,
and are approaching SNR levels of nearly single spike accuracy as well
[\citet{WallaceHasan08}]. Microscopy technologies for collecting
fluorescence signals are also rapidly developing. Cooled CCDs for
wide-field imaging (either epifluorescence or confocal) now achieve a
quantum efficiency of $\approx$90$\%$ with frame rates up to $60$ Hz
or greater, depending on the field of view [\citet{Djurisic04}]. For in
vivo work, 2-photon laser scanning microscopy can achieve similar
frame rates, using either acoustic-optical deflectors to focus light
at arbitrary locations in three-dimensional space
[\citet{Iyer06}; \citet{SalomeBourdieu06}; \citet{ReddySaggau08}] or resonant scanners
[\citet{NguyenParker01}]. Together, these experimental tools can provide
movies of calcium fluorescence transients from large networks of
neurons with adequate SNR, at imaging frequencies of $30$ Hz or
greater, in both {in vitro} and {in vivo} preparations.

\begin{figure}[b]

\includegraphics{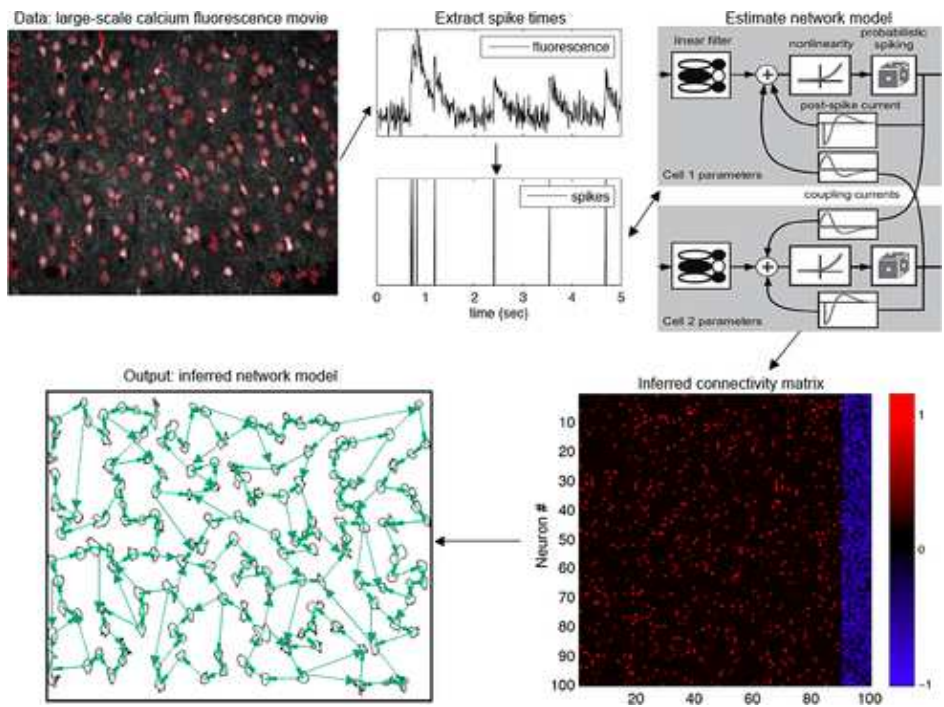}

\caption{Schematic overview. The raw observed data is a
large-scale calcium fluorescence movie, which is pre-processed
to correct for movement artifacts and find
regions-of-interest, that is, putative neurons. [Note that we
have omitted details of these important preprocessing steps in
this paper; see, for example,  Cossart, Aronov and Yuste (\protect\citeyear{CAR03});
 Dombeck et~al. (\protect\citeyear{DombeckTank07}) for further
details.] Given the fluorescence traces $F_i(t)$ from each
neuron, we estimate the underlying spike trains (i.e., the
time series of neural activity) using statistical
deconvolution methods. Then we estimate the parameters of a
network model given the observed data. Our major goal is to
obtain an accurate estimate of the network connectivity
matrix, which summarizes the information we are able to infer
about the local neuronal microcircuit. (We emphasize that
this illustration is strictly schematic, and does not
correspond directly to any of the results described below.)
This figure adapted from personal communications with R.
Yuste, B. Watson and A. Packer.}
\label{fig:data_schematic}
\end{figure}

Given these experimental advances in functional neural imaging, our
goal is to develop efficient computational and statistical methods to
exploit this data for the analysis of neural connectivity; see Figure
\ref{fig:data_schematic} for a schematic overview. One major challenge
here is that calcium transients due to action potentials provide
indirect observations, and decay about an order of magnitude slower
than the time course of the underlying neural activity
[\citet{ImagingManual}; \citet{Roxin08}]. Thus, to properly analyze the network
connectivity, we must incorporate methods for effectively deconvolving
the observed noisy fluorescence signal to obtain estimates of the
underlying spiking rates
[\citet{YaksiFriedrich06}; \citet{GreenbergKerr08}; \citet{Vogelstein2009}]. To this end, we
introduce a coupled Markovian state-space model that relates the
observed variables (fluorescence traces from the neurons in the
microscope's field of view) to the hidden variables of interest (the
spike trains and intracellular calcium concentrations of these
neurons), as governed by a set of biophysical parameters including the
network connectivity matrix. As discussed in [\citet{Vogelstein2009}],
this parametric approach effectively introduces a number of
constraints on the hidden variables, leading to significantly better
performance than standard blind deconvolution approaches. Given this
state-space model, we derive a Monte Carlo Expectation--Maximization
algorithm for obtaining the maximum a posteriori estimates of the
parameters of interest. Standard sampling procedures (e.g., Gibbs
sampling or sequential Monte Carlo) are inadequate in this setting,
due to the high dimensionality and nonlinear, non-Gaussian dynamics
of the hidden variables; we therefore develop a specialized
blockwise-Gibbs approach for efficiently computing the sufficient
statistics. This strategy enables us to accurately infer the
connectivity matrix from large simulated neural populations, under
realistic assumptions about the dynamics and observation parameters.

\section{Methods}
\label{sec:methods}
\subsection{Model}
\label{sec:methods:markov-setup}

We begin by detailing a parametric generative model for the
(unobserved) joint spike trains of all $N$ observable neurons, along
with the observed calcium fluorescence data. Each neuron is modeled as
a generalized linear model (GLM). This class of models is known to
capture the statistical firing properties of individual neurons fairly
accurately
[\citet{BRIL88}; \citet{CSK88}; \citet{BRIL92}; \citet{PG00}; \citet{PAN03d};
\citet{PAN04c}; \citet{Rigat06}; \citet{TRUC05}; \citet{NYK06}; \citet{KP06}; \citet{PILL07}; \citet{Vidne08}; \citet{Stevenson2009}].
We
denote the $i$th neuron's activity at time $t$ as $n_i(t)$: in
continuous time, $n_i(t)$ could be modeled as an unmarked point
process, but we will take a~discrete-time approach here, with each
$n_i(t)$ taken to be a binary random variable. We model the spiking
probability of neuron $i$ via an instantaneous nonlinear function,
$f(\cdot)$, of the filtered and summed input to that neuron at that
time,~$J_i(t)$. This input is composed of the following: (i) some
baseline value,~$b_i$; (ii)~some external vector stimulus, $S^\mathit{ext}(t)$, that is
linearly filtered by~$k_i$; and (iii) spike history terms,
$h_{ij}(t)$, encoding the influence on neuron~$i$ from neuron $j$,
weighted by $w_{ij}$:
%
%
\begin{equation} \label{eqn:glm:definition}
\hspace*{15pt}n_i(t) \sim\operatorname{Bernoulli}[f(J_i(t) ) ],\qquad
J_i(t)=b_i+k_i\cdot S^\mathit{ext}(t)+\sum _{j=1}^{N} w_{ij} h_{ij}(t).
\end{equation}

\begin{figure}[b]

\includegraphics{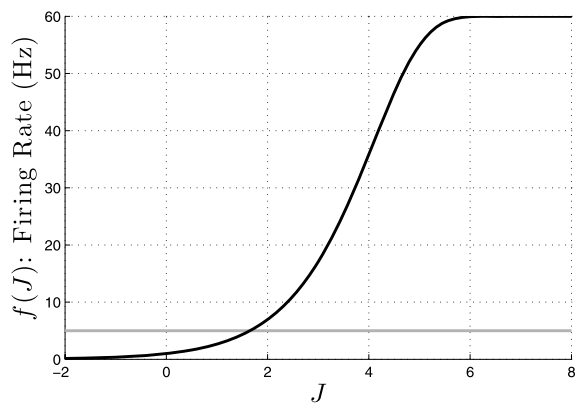}

\caption{A plot of the firing rate nonlinearity $f(J)$ used in our
simulations. Note that the firing rate saturates at $1/\Delta$,
because of our Bernoulli assumption (i.e., the spike count per bin
is at most one). Here the binwidth $\Delta= (60$ Hz$)^{-1}$. The
horizontal gray line indicates~5~Hz, the baseline firing rate for
most of the simulations discussed in the Results section.}
\label{fig:egfluor}
\end{figure}

To ensure computational tractability of the parameter inference
problem, we must impose some reasonable constraints on the
instantaneous nonlinearity $f(\cdot)$ (which plays the role of the
inverse of the link function in the standard GLM setting) and on the
dynamics of the spike-history effects~$h_{ij}(t)$. First, we restrict
our attention to functions $f(\cdot)$ which ensure the concavity of
the spiking loglikelihood in this model [\citet{PAN04c}; \citet{Escola07}], as we
will discuss at more length below. In this paper we use\looseness=-1
%
%
\begin{equation}
f(J) = P[n>0  \mid  n \sim\operatorname{Poiss}(e^J\Delta
)]
= 1 - \exp[-e^J \Delta]
\end{equation}
(Figure \ref{fig:egfluor}), where the inclusion of $\Delta$, the time
step size, ensures that the firing rate scales properly with respect
to the time discretization; see [\citet{Escola07}] for a proof that this
$f(\cdot)$ satisfies the required concavity constraints. However, we
should note that in our experience the results depend only weakly on
the details of $f(\cdot)$ within the class of log-concave models
[\citet{LD89}; \citet{PAN04c}] (see also Section \ref{sec:impact-strong-corr}
below).

Second, because the algorithms we develop below assume Markovian
dynamics, we model the spike history terms as autoregressive
processes driven by the spike train $n_j(t)$:
%
%
\begin{equation} \label{eqn:h:definition}
h_{ij}(t) = (1- \Delta/\tau^h_{ij}) h_{ij}(t- \Delta) +n_j(t-\Delta
) +
\sigma^h_{ij} \sqrt{\Delta} \varepsilon^h_{ij}(t),
\end{equation}
where $\tau^h_{ij}$ is a decay time constant, $\sigma^h_{ij}$ is a
standard deviation parameter, $\sqrt{\Delta}$ ensures that the
statistics of this Markov process have a proper Ornstein--Uhlenbeck
limit as $\Delta\to0$, and throughout this paper, $\varepsilon$ denotes
an independent standard normal random variable. Note that this model
generalizes [via a simple augmentation of the state variable
$h_{ij}(t)$] to allow each neuron pair to have several spike history
terms, each with a unique time constant, which when weighted and
summed allow us to model a wide variety of possible post-synaptic
effects, including bursting, facilitating, and depressing synapses;
see [\citet{Vogelstein2009}] for further details.\vspace*{1pt} We restrict our
attention to the case of a single time constant $\tau^h_{ij}$ per
synapse here, so the deterministic part of $h_{ij}(t)$ is a simple
exponentially-filtered version of the spike train
$n_j(t)$. Furthermore, we assume that $\tau^h_{ij}$ is the same for
all neurons and all synapses, although, in principle, each synapse could
be modeled with its unique $\tau^h_{ij}$. We do that both for
simplicity and also because we find that the detailed shape of the
coupling terms $h_{ij}(t)$ had a limited effect on the inference of
the connectivity matrix, as illustrated in Figure \ref{fig:vartau}
below. Thus, we treat $\tau^h_{ij}$ and $\sigma^h_{ij}$ as known
synaptic parameters which are the same for each neuron pair $(i,j)$,
and denote them as $\tau_h$ and $\sigma_h$ hereafter. We chose values
for $\tau_h$ and $\sigma_h$ in our inference based on experimental
data [\citet{Lefort2009}]; see Table 1 below. Therefore, our unknown
spiking parameters are $\{\mathbf{w}_i,k_i,b_i\}_{i\leq N}$, with
$\mathbf{w}_i=(w_{i1},\ldots, w_{iN})$.

\begin{figure}[b]

\includegraphics{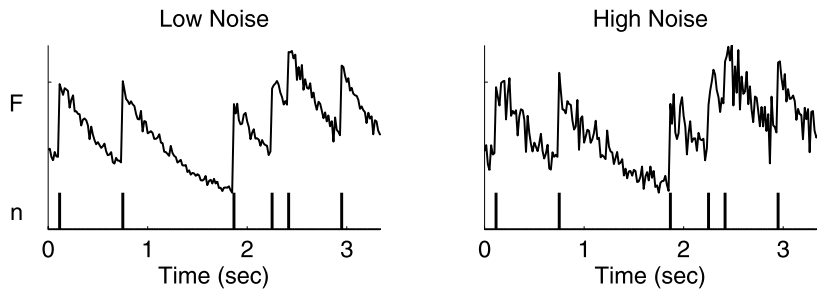}

\caption{Two example traces of simulated fluorescence data, at
different SNR levels, demonstrating the relationship between spike
trains and observed fluorescence in our model. Note that both
panels have the same underlying spike train. Simulation parameters:
$k_i=0.7$, $C_i^b=1$ $\mu$M, $\tau^c_i=500$ msec, $A_i=50$ $\mu$M,
$\sigma^c_i=0.1$ $\mu$M, $\gamma_i=0.004$ [effective SNR $\approx
9$, as defined in equation \textup{(\protect\ref{eq:eSNR})}; see also
Figure~\textup{\protect\ref{fig:recvar-SNR}} below] in the left panel and $\gamma_i=0.016$
(eSNR $\approx3$) in the right panel, and $\sigma^F_i=0$,
$\Delta=(60$~Hz$)^{-1}$.}
\label{fig:example_traces}
\end{figure}

The problem of estimating the connectivity parameters
$\mathbf{w}=\{\mathbf{w}_i\}_{i\leq N}$ in this type of GLM, given a
fully-observed
ensemble of neural spike trains $\{n_i(t)\}_{i\leq N}$, has recently
received a great deal of attention; see the references above for a
partial list. In the calcium fluorescent imaging setting, however, we
do not directly observe spike trains; $\{n_i(t)\}_{i\leq N}$ must be
considered a~hidden variable here. Instead, each spike in a given
neuron leads to a rapid increase in the intracellular calcium
concentration, which then decays slowly due to various cellular
buffering and extrusion mechanisms. We in turn make only noisy,
indirect, and subsampled observations of this intracellular calcium
concentration, via fluorescent imaging techniques
[\citet{ImagingManual}]. To perform statistical inference in this
setting, [\citet{Vogelstein2009}] proposed a simple conditional
first-order hidden Markov model (HMM) for the intracellular calcium
concentration $C_i(t)$ in cell $i$ at time $t$, along with the
observed fluorescence, $F_i(t)$:
%
\begin{eqnarray}
\label{eqn:ca:definition}
C_i(t) &=& C_i(t-\Delta) + \bigl( C_i^b-C_i(t-\Delta) \bigr) \Delta/
\tau^c_i + A_i n_i(t) + \sigma^c_i \sqrt{\Delta} \varepsilon^c_i(t),
\\
F_i(t) &=& \alpha_i S(C_i(t)) + \beta_i + \sqrt{(\sigma^F_i)^2 +
\gamma_i S(C_i(t)) } \varepsilon^F_i(t). \label{eqn:F:definition}
\end{eqnarray}
This model can be interpreted as a simple driven autoregressive
process: under nonspiking conditions, $C_i(t)$ fluctuates around the
baseline level of $C_i^b$, driven by normally-distributed noise
$\varepsilon^c_i(t)$ with standard deviation $\sigma^c_i
\sqrt{\Delta}$. Whenever the neuron fires a spike, $n_i(t)=1$, the
calcium variable $C_i(t)$ jumps by a fixed amount~$A_i$, and
subsequently decays with time constant $\tau^c_i$. The fluorescence
signal $F_i(t)$ corresponds to the count of photons collected at the
detector per neuron per imaging frame. This photon count may be
modeled with normal statistics, with the mean given by a saturating
Hill-type function $S(C)=C/(C+K_d)$ [\citet{Yasuda2004}] and the variance
scaling with the mean; see [\citet{Vogelstein2009}] for further
discussion. Because the parameter $K_d$ effectively acts as a simple
scale factor, and is a~property of the fluorescent indicator, we
assume throughout this work that it is known. Figure
\ref{fig:example_traces} shows a couple examples depicting the
relationship between spike trains and observations. It will be useful
to define an effective SNR as
%
%
\begin{equation}
\operatorname{eSNR} = \frac{E[F_i(t)-F_i(t-\Delta)  |  n_i(t)=1]}
{E[(F_i(t)-F_i(t-\Delta))^2/2  |  n_i(t)=0]^{1/2}},
\label{eq:eSNR}
\end{equation}
that is, the size of a spike-driven fluorescence jump divided by a rough
measure of the standard deviation of the baseline fluorescence. For
concreteness, the effective SNR values in
Figure \ref{fig:example_traces} were $9$ and $3$ in the left and right
panels, respectively.

To summarize, equations (\ref{eqn:glm:definition})--(\ref{eqn:F:definition})
define a coupled HMM: the underlying spike trains $\{n_i(t)\}_{i\leq
N}$ and spike history terms $\{h_{ij}(t)\}_{i,j\leq N}$\vadjust{\goodbreak} evolve in
a~Markovian manner given the stimulus $S^\mathit{ext}(t)$. These spike trains
in turn drive the intracellular calcium concentrations
$\{C_i(t)\}_{i\leq N}$, which are themselves Markovian, but evolving
at a slower timescale $\tau_i^c$. Finally, we observe only the
fluorescence signals $\{F_i(t)\}_{i\leq N}$, which are related in a
simple Markovian fashion to the calcium variables $\{C_i(t)\}_{i\leq
N}$.

\subsection{Goal and general strategy} \label{sec:methods:goal}

Our primary goal is to estimate the connectivity matrix, $\mathbf{w}$, given
the observed set of calcium fluorescence signals $\mathbf{F}=\{\mathbf
{F}_i\}_{i\leq
N}$, where $\mathbf{F}_i=\{F_i(t)\}_{t\leq T}$. We must also deal with a
number of intrinsic parameters,\footnote{The intrinsic parameters for
neuron $i$ are all its parameters minus the cross-coupling terms,
that is,\ $\tilde{\mathbf{\theta}}_i =\mathbf{\theta}_i \backslash
\{w_{ij}\}_{i\neq j}$.} $\tilde{\mathbf{\theta}}_i$:
the intrinsic spiking parameters\footnote{To reduce the notational
load, we will ignore the estimation of the stimulus filter $k_i$
below; this term may be estimated with $b_i$ and $w_{ii}$ using very
similar convex optimization methods, as discussed in
[\citet{Vogelstein2009}].} $\{b_i, w_{ii}\}_{i\leq N}$, the calcium
parameters $\{C^b_i, \tau^c_i, A_i, \sigma^c_i\}_{i\leq N}$, and the
observation parameters $\{\alpha_i, \beta_i, \gamma_i,
\sigma^F_i\}_{i\leq N}$. We addressed the problem of estimating these
intrinsic parameters in earlier work [\citet{Vogelstein2009}]; thus, our
focus here will be on the connectivity matrix $\mathbf{w}$. A Bayesian
approach is natural here, since we have a good deal of prior
information about neural connectivity; see [\citet{Rigat06}] for a
related discussion. However, a fully-Bayesian approach, in which we
numerically integrate over the very high-dimensional parameter space
$\mathbf{\theta}= \{\mathbf{\theta}_i\}_{i\leq N}$, where $\mathbf
{\theta}_i=\{\mathbf{w}_i, b_i, C^b_i,
\tau^c_i, A_i, \sigma^c_i, \alpha_i, \beta_i, \gamma_i, \sigma
^F_i\}$,
is less attractive from a computational point of view. Thus, our
compromise is to compute \textit{maximum a posteriori} (MAP) estimates
for the parameters via an expectation--maximization (EM) algorithm, in
which the sufficient statistics are computed by a hybrid blockwise
Gibbs sampler and sequential Monte Carlo (SMC) method. More
specifically, we iterate the steps:
\begin{longlist}[E step:]
\item[E-step:] Evaluate  $Q(\mathbf{\theta},\mathbf
{\theta}^{(l)}) = E_{P[\mathbf{X}|
\mathbf{F}; \mathbf{\theta}^{(l)}]} \ln P [ \mathbf{F},
\mathbf{X}| \mathbf{\theta}] =
 \int P[\mathbf{X}|
\mathbf{F}; \mathbf{\theta}^{(l)}] \ln P[\mathbf{F}\hspace*{-0,4pt},\\\hspace*{48pt} \mathbf{X}|
\mathbf{\theta}]\, d \mathbf{X}$;\vspace*{3pt}
\item[M-step:] Solve  $\mathbf{\theta}^{(l+1)} = \operatorname
{argmax}_{\mathbf{\theta}} \{
Q(\mathbf{\theta},\mathbf{\theta}^{(l)}) + \ln P(\mathbf{\theta})
\}$,
\end{longlist}
where $\mathbf{X}$ denotes the set of all hidden variables $\{ C_i(t), n_i(t),
h_{ij}(t) \}_{i,j \leq N, t \leq T}$ and~$P(\mathbf{\theta})$ denotes
a (possibly
improper) prior on the parameter space $\mathbf{\theta}$. According
to standard EM
theory [\citet{DLR77}; \citet{McLachlanKrishnan96}], each iteration of these two
steps is guaranteed to increase the log-posterior $\ln P(\mathbf
{\theta}^{(l)} |
\mathbf{F})$, and will therefore lead to at least a~locally maximum a
posteriori estimator.

Now, our major challenge is to evaluate the auxiliary function
$Q(\mathbf{\theta},\mathbf{\theta}^{(l)})$ in the E-step. Our model
is a coupled HMM, as
discussed in the previous section; therefore, as usual in the HMM
setting [\citet{RAB89}], $Q$ may be broken up into a sum of simpler
terms:
%
%
\begin{eqnarray}\label{eqn:loglik:definition-expl}
Q\bigl(\mathbf{\theta},\mathbf{\theta}^{(l)}\bigr) &=& \sum
_{it} \int\ln
P[F_i(t)|C_i(t); \alpha_i, \beta_i, \gamma_i, \sigma^F_i] \,dP\bigl[C_i(t)
| \mathbf{F}; \theta^{(l)}\bigr] \nonumber\\
&&{}+ \sum_{it} \int\ln P[C_i(t) |
C_i(t-\Delta),
\nonumber
\\[-8pt]
\\[-8pt]
\nonumber
&&\qquad{}\hspace*{41pt} n_i(t); C^b_i, \tau^c_i, A_i, \sigma^c_i] \,dP\bigl[C_i(t),
C_i(t-\Delta) | \mathbf{F}; \theta^{(l)}\bigr] \\
&&{}+ \sum
_{it} \int
\ln P[n_i(t)| \mathbf{h}_i(t); b_i, \mathbf{w}_i]\, dP\bigl[n_i(t), \mathbf
{h}_i(t) | \mathbf{F};
\mathbf{\theta}^{(l)}\bigr],\nonumber
\end{eqnarray}
where $\mathbf{h}_i(t)=\{h_{ij}(t)\}_{j \leq N}$. Note that each of
the three
sums here corresponds to a different component of the model described
in equations (\ref{eqn:glm:definition})--(\ref{eqn:F:definition}): the first
sum involves the fluorescent observation parameters, the second the
calcium dynamics, and the third the spiking dynamics.

Thus, we need only compute low-dimensional marginals of the full
posterior distribution $P[\mathbf{X}| \mathbf{F}; \mathbf{\theta
}]$; specifically, we need the
pairwise marginals $P[C_i(t)| \mathbf{F}; \mathbf{\theta}]$,
$P[C_i(t), C_i(t- \Delta) |
\mathbf{F}; \mathbf{\theta}]$, and $P[n_i(t),\mathbf{h}_i(t) |
\mathbf{F}; \mathbf{\theta}]$. Details for
calculating $P[C_i(t), C_i(t- \Delta) | \mathbf{F}_i; \tilde{\mathbf
{\theta}}_i]$ and $P[C_i(t)
| \mathbf{F}_i; \tilde{\mathbf{\theta}}_i]$ are found in [\citet
{Vogelstein2009}], while
calculating the joint marginal for the high-dimensional hidden
variable $\mathbf{h}_i$ necessitates the development of specialized blockwise
Gibbs-SMC sampling methods, as we describe in the subsequent Sections~%
\ref{sec:methods:indep} and \ref{sec:methods:joint}. Once we have
obtained these marginals, the M-step breaks up into a~number of
independent optimizations that may be computed in parallel and which
are therefore relatively straightforward (Section
\ref{sec:methods:parametersHMM}); see Section~%
\ref{sec:methods:specific_implementation} for a pseudocode summary
along with some specific implementation details.

\subsection{Initialization of intrinsic parameters via sequential
Monte Carlo methods} \label{sec:methods:indep}

We begin by constructing relatively cheap, approximate preliminary
estimators for the intrinsic parameters, $\tilde{\mathbf{\theta
}}_i$. The idea is to
initialize our estimator by assuming that each neuron is observed
independently. Thus, we want to compute $P[C_i(t), C_i(t-\Delta) |
\mathbf{F}_i; \tilde{\mathbf{\theta}}_i]$ and $P[C_i(t)|\mathbf
{F}_i;\tilde{\mathbf{\theta}}_i]$, and solve the M-step
for each~$\tilde{\mathbf{\theta}}_i$, with the connectivity matrix
parameters held
fixed. This single-neuron case is much simpler, and has been discussed
at length in [\citet{Vogelstein2009}]; therefore, we only provide a brief
overview here. The standard forward and backward recursions provide
the necessary posterior distributions, in principle
[\citet{ShumwayStoffer06}]:
%
\begin{eqnarray}
&&P[X_i(t) | F_i(0\dvtx t)] \nonumber\\
&&\qquad\propto P[F_i(t)| X_i(t)] \int P[X_i(t) |
X_i(t-\Delta)]\label{eqn:forward}\\
&&\qquad{}\hspace*{88pt}\times P[X_i(t-\Delta) | F_i(0\dvtx t-\Delta)] \, dX_i(t-\Delta),\nonumber\\
&&P[X_i(t), X_i(t-\Delta) | \mathbf{F}_i] \nonumber\\
&&\qquad= P[X_i(t) | \mathbf{F}_i]\label{eqn:backward}\\
&&\qquad{}\hspace*{8pt}\times\frac{P[X_i(t) | X_i(t-\Delta)] P[X_i(t-\Delta) |
F_i(0\dvtx t-\Delta)]}{\int P[X_i(t) | X_i(t-\Delta)] P[X_i(t-\Delta) |
F_i(0\dvtx t-\Delta)]\, dX_i(t-\Delta)},\nonumber
\end{eqnarray}
where $F_i(s\dvtx t)$ denotes the time series $\mathbf{F}_i$ from time points
$s$ to $t$, and we have dropped the conditioning on the parameters for
brevity's sake. Equation~\eqref{eqn:forward} describes the forward (filter)
pass of the recursion, and equation \eqref{eqn:backward} describes the
backward (smoother) pass, providing both $P[X_i(t), X_i(t-\Delta) |
\mathbf{F}_i]$ and $P[X_i(t) | \mathbf{F}_i]$ [obtained by
marginalizing over
$X_i(t-\Delta)$].

Because these integrals cannot be analytically evaluated for our
model, we approximate them using a SMC (``marginal particle
filtering'') method [\citet{DGA00}; \citet{DFG01}; \citet{GDW04}]. More specifically, we
replace the forward distribution with a particle approximation:
%
\begin{equation}\label{eqn:forward2}
P[X_i(t) | F_i(0\dvtx t)] \approx\sum_{m=1}^M p_f^{(m)}(t) \delta
\bigl[X_i(t)-X_i^{(m)}(t)\bigr],
\end{equation}
where $m=1,\ldots,M$ indexes the $M$ particles in the set ($M$ was
typically set to about $50$ in our experiments), $p_f^{(m)}(t)$
corresponds to the relative ``forward''\vspace*{-2pt} probability of
$X_i(t)=X_i^{(m)}(t)$, and $\delta[\cdot]$ indicates a Dirac
mass. Instead of using the analytic forward recursion,
equation \eqref{eqn:forward}, at each time step, we update the particle
weights using the particle forward recursion
%
\begin{equation}
p_f^{(m)}(t) = P\bigl[F_i(t) | X_i^{(m)}(t)\bigr]
\frac{P[X_i^{(m)}(t) | X_i^{(m)}(t-\Delta)]
p_f^{(m)}(t-\Delta)}{q[X_i^{(m)}(t)]},
\label{eq:forward-particle}
\end{equation}
where $q[X_i^{(m)}(t)]$ is the proposal density from which we
sample the particle positions $X_i^{(m)}(t)$. In this work we use
the ``one-step-ahead'' sampler [\citet{DGA00}; \citet{Vogelstein2009}], that is,
$q[X_i^{(m)}(t)]=P[X_i^{(m)}(t) | X_i^{(m)}(t-\Delta),
F_i(t)]$. After sampling and computing the weights, we use
stratified resampling [\citet{DCM05}] to ensure the particles accurately
approximate the desired distribution. Once we complete the forward
recursion from $t=0,\ldots, T$, we begin the backward pass from
$t=T,\ldots, 0$, using
%
\begin{eqnarray}
\hspace*{14pt}r^{(m,m')}(t,t-\Delta) &=& p_b^{(m)}(t) \frac{P[X_i^{(m)}(t) |
X_i^{(m')} (t-\Delta)] p_f^{(m)}(t-\Delta)} {\sum_{m'}
P[X_i^{(m)}(t) | X_i^{(m')}(t-\Delta)] p_f^{(m')} (t-\Delta)},
\\ p_b^{(m')}(t-\Delta) &=& \sum_{j=1}^M r^{(m,m')} (t,t-\Delta),
\label{eqn:backward2}
\end{eqnarray}
to obtain the approximation
%
\begin{eqnarray}\label{eqn:particle-fb}
&&P[X_i(t), X_i(t-\Delta) | F_i] \nonumber\\
&&\qquad\approx\sum_{m,m'}
r_i^{(m,m')}(t,t-\Delta) \delta\bigl[ X_i(t) - X_i^{(m)}(t) \bigr]\\
&&\qquad{}\hspace*{27pt}\times\delta\bigl[ X_i(t-\Delta) - X_i^{(m')}(t-\Delta)
\bigr]\nonumber
\end{eqnarray}
for more details, see [\citet{Vogelstein2009}]. Thus, equations
(\ref{eqn:forward2})--(\ref{eqn:particle-fb}) may be used to compute the
sufficient statistics for estimating the intrinsic parameters
$\tilde{\mathbf{\theta}}_i$ for each neuron.

As discussed following equation \eqref{eqn:loglik:definition-expl}, the
M-step decouples into three independent subproblems. The first term
depends on only $\{\alpha_i, \beta_i, \gamma_i, \sigma_i\}$; since
$P[F_i(t)|S(C_i(t)); \tilde{\mathbf{\theta}}_i]$ is Gaussian, we can
estimate these
parameters by solving a weighted regression problem (specifically, we
use a coordinate--optimi\-zation approach: we solve a quadratic problem
for $\{\alpha_i, \beta_i\}$ while holding $\{\gamma_i, \sigma_i\}$
fixed, then estimate $\{\gamma_i,\sigma_i\}$ by the usual residual
error formulas while holding $\{\alpha_i, \beta_i\}$
fixed). Similarly, the second term requires us to optimize over
$\{\tau_i^c, A_i, C_i^b\}$, and then we use the residuals to estimate
$\sigma_i^c$. Note that all the parameters mentioned so far are
constrained to be non-negative, but may be solved efficiently using
standard quadratic program solvers if we use the simple
reparameterization $\tau_i^c \to1- \Delta/ \tau_i^c$. Finally, the
last term may be expanded:
%
\begin{eqnarray}\label{eqn:bw}
&&E [\ln P[n_i(t), \mathbf{h}_i(t) | \mathbf{F};
\mathbf{\theta}_i]]\nonumber
\\
&&\qquad= P[n_i(t), \mathbf{h}_i(t) |
\mathbf{F}; \mathbf{\theta}_i] \ln f [J_i(t)] \\
&&\quad \qquad{}+ \bigl(1-P[n_i(t), \mathbf
{h}_i(t) | \mathbf{F}; \mathbf{\theta}_i]\bigr)
\ln[1- f(J_i(t))];\nonumber
\end{eqnarray}
since $J_i(t)$ is a linear function of $\{b_i,\mathbf{w}_i\}$, and the
right-hand side of equation~\eqref{eqn:bw} is concave in $J_i(t)$, we see
that the third term in equation \eqref{eqn:loglik:definition-expl} is
a~sum of terms which are concave in $\{b_i,\mathbf{w}_i\}$---and therefore also
concave in the linear subspace $\{b_i,w_{ii}\}$ with $\{w_{ij}\}_{i
\neq j}$ held fixed---and may thus be maximized efficiently using
any convex optimization method, for example,\ Newton--Raphson or conjugate
gradient ascent.

Our procedure therefore is to initialize the parameters for each
neuron using some default values that we have found to be effective in
practice in analyzing real data, and then iteratively (i) estimate the
marginal posteriors via the SMC recursions
(\ref{eqn:forward2})--(\ref{eqn:particle-fb}) \mbox{(E-step)}, and (ii) maximize
over the intrinsic parameters~$\tilde{\mathbf{\theta}}_i$ \mbox{(M-step)},
using the separable
convex optimization approach described above. We iterate these two
steps until the change in $\tilde{\mathbf{\theta}}_i$ does not
exceed some minimum
threshold. We then use the marginal posteriors from the last iteration
to seed the blockwise Gibbs sampling procedure described below for
approximating $P[n_i,\mathbf{h}_i | \mathbf{F};\mathbf{\theta}_i]$.

\subsection{Estimating joint posteriors over weakly coupled neurons}
\label{sec:methods:joint}

Now we turn to the key problem: constructing an estimate of the joint
marginals $\{P[n_i(t), \mathbf{h}_i(t) | \mathbf{F}; \mathbf{\theta
}]\}_{i\leq N,
t\leq T}$, which are the sufficient statistics for estimating the
connectivity matrix $\mathbf{w}$ [recall
equation \eqref{eqn:loglik:definition-expl}]. The SMC method described in
the preceding section only provides the marginal distribution over a
single neuron's hidden \mbox{variables}; this method may in principle be
extended to obtain the desired full posterior $P[\mathbf{X}(t),
\mathbf{X}(t-\Delta)
| \mathbf{F}; \mathbf{\theta}]$, but SMC is fundamentally a
sequential importance
sampling method, and therefore scales poorly as the dimensionality of
the hidden state~$\mathbf{X}(t)$ increases [\citet{BickelBengtsson08}].
Thus, we
need a different approach.

One very simple idea is to use a Gibbs sampler: sample sequentially
from
%
\begin{equation}
\hspace*{24pt}X_i(t) \sim P[X_i(t) | \mathbf{X}_{\backslash i}, X_i(0), \ldots,
X_i(t-\Delta),
X_i(t+\Delta), \ldots, X_i(T), \mathbf{F}; \mathbf{\theta}],
\end{equation}
looping over all cells $i$ and all time bins $t$. Unfortunately, this
approach is likely to mix poorly, due to the strong temporal dependence
between $X_i(t)$ and $X_i(t+\Delta)$. Instead, we propose a blockwise
Gibbs strategy, sampling one spike train as a block:
%
\begin{equation}
\mathbf{X}_i \sim P[\mathbf{X}_i | \mathbf{X}_{\backslash i},
\mathbf{F}; \mathbf{\theta}].
\end{equation}
If we can draw these blockwise samples $\mathbf{X}_i = \mathbf{X}_i(s\dvtx t)$
efficiently for a large subset of $t-s$ adjacent time-bins
simultaneously, then we would expect the resulting Markov chain to mix
much more quickly than the single-element Gibbs chain. This follows
due to the weak dependence between $\mathbf{X}_i$ and $\mathbf{X}_j$
when $i\neq j$,
and the fact that Gibbs is most efficient for weakly-dependent
variables [\citet{RC05}].

So, how can we efficiently sample from $P[\mathbf{X}_i | \mathbf
{X}_{\backslash i}, \mathbf{F};
\mathbf{\theta}]$? One attractive approach is to try to re-purpose
the SMC method
described above, which is quite effective for drawing approximate
samples from $P[\mathbf{X}_i | \mathbf{X}_{\backslash i}, F_i;
\mathbf{\theta}]$ for one neuron $i$ at a
time. Recall that sampling from an HMM is in principle easy by the
``propagate forward, sample backward'' method: we first compute the
forward probabilities $P[X_i(t) | \mathbf{X}_{\backslash i}(0\dvtx t),
F_i(0\dvtx t); \mathbf{\theta}]$
recursively for timesteps $t=0$ up to $T$, then sample backward from
$P[X_i(t) | \mathbf{X}_{\backslash i}(0\dvtx T), F_i(0\dvtx T), X_i(t-\Delta);
\mathbf{\theta}]$. This
approach is powerful because each sample requires just linear time to
compute [i.e., $O(T/\Delta)$ time, where $T/\Delta$ is the number of
desired time steps]. Unfortunately, in this case we can only compute
the forward probabilities approximately (via
equations \mbox{(\ref{eqn:forward2})--(\ref{eq:forward-particle}))}, and so therefore
this attractive forward-backward approach only provides approximate
samples from $P[\mathbf{X}_i | \mathbf{X}_{\backslash i}, \mathbf
{F}; \mathbf{\theta}]$, not the exact samples
required for the validity of the Gibbs method.

Of course, in principle, we should be able to use the
Metropolis--Hastings \mbox{(M--H)} algorithm to correct these approximate
samples. The problem is that the M--H acceptance ratio in this setting
involves a high-dimensional integral over the set of paths that the
particle filter might possibly trace out, and is therefore difficult
to compute directly. [\citet{Andrieu2007}] discuss this problem at more
length, along with some proposed solutions. A slightly simpler
approach was introduced by [\citet{NBR03}]. Their idea is to exploit the
$O(T/\Delta)$ forward-backward sampling method by embedding a discrete
Markov chain within the continuous state space $\mathcal{X}_t$ on
which $X_i(t)$ is defined; the state space of this discrete embedded
chain is sampled randomly according to some distribution $\rho_t$ with
support on $\mathcal{X}_t$. It turns out that an appropriate Markov
chain (incorporating the original state space model transition and
observation probabilities, along with the auxiliary sampling
distributions $\rho_t$) may be constructed quite tractably,
guaranteeing that the samples produced by this algorithm have the
desired equilibrium density. See [\citet{NBR03}] for
details.\looseness=-1

We can apply this embedded-chain method directly here to sample from
$P[\mathbf{X}_i | \mathbf{X}_{\backslash i}, \mathbf{F}; \mathbf
{\theta}]$. The one remaining question is how to
choose the auxiliary densities $\rho_t$. We would like to choose these
densities to be close to the desired marginal densities $P[X_i(t) |
\mathbf{X}_{\backslash i}, \mathbf{F}; \mathbf{\theta}]$, and
conveniently, we have already computed a
good (discrete) approximation to these densities, using the SMC
methods described in the last section. The algorithm described in
[\citet{NBR03}] requires the densities~$\rho_t$ to be continuous, so we
simply convolve our discrete SMC-based approximation [specifically,
the $X_i(t)$-marginal of equation (\ref{eqn:particle-fb})] with an
appropriate normal density to arrive at a very tractable
mixture-of-Gaussians representation for $\rho_t$.

Thus, to summarize, our procedure for approximating the desired joint
state distributions $P[n_i(t), \mathbf{h}_i(t) | \mathbf{F};\mathbf
{\theta}]$ has a
Metropolis-within-blockwise-\break Gibbs flavor, where the internal
Metropolis step is replaced by the $O(T/\Delta)$ embedded-chain method
introduced by [\citet{NBR03}], and the auxiliary densities~$\rho_t$
necessary for implementing the embedded-chain sampler are obtained
using the SMC methods from [\citet{Vogelstein2009}].

\subsubsection{A factorized approximation of the joint posteriors}
\label{sec:cheaper-high-snr}

If the SNR in the calcium imaging is sufficiently high, then, by
definition, the observed fluorescence data $F_i$ will provide enough
information to determine the underlying hidden variables
$\mathbf{X}_i$. Thus, in this case the joint posterior approximately
factorizes into a product of marginals for each neuron $i$:
%
%
\begin{equation} \label{eqn:indep_approx}
P[\mathbf{X}|\mathbf{F};\mathbf{\theta}] \approx\prod_{i\leq N}
P[\mathbf{X}_i | \mathbf{F}; \tilde{\mathbf{\theta}}_i].
\end{equation}
We can take advantage of this because we have already estimated all
the marginals on the right-hand side using the approximate SMC methods
in Section \ref{sec:methods:indep}. This factorized approximation
entails a significant gain in efficiency for two reasons: first, it
obviates the need to generate joint samples via the expensive
blockwise-Gibbs approach described above; and second, because we can
easily parallelize the SMC step, inferring the marginals $P[X_i(t) |
F_i; \tilde{\mathbf{\theta}}_i]$ and estimating the parameters~$\mathbf{\theta}_i$ for each neuron
on a separate processor. We will discuss the empirical accuracy of
this approximation in Section \ref{sec:results}.

\subsection{Estimating the connectivity matrix} \label{sec:methods:parametersHMM}

Computing the M-step for the connectivity matrix, $\mathbf{w}$, is an
optimization problem with on the order of $N^2$ variables. The
auxiliary function equation \eqref{eqn:loglik:definition-expl} is
concave in $\mathbf{w}$, and decomposes into $N$ separable terms that
may be
optimized independently using standard ascent methods. To improve our
estimates, we will incorporate two sources of strong {a priori}
information via our prior $P(\mathbf{w})$: first, previous anatomical studies
have established that connectivity in many neuroanatomical substrates
is ``sparse,'' that is, most neurons form synapses with only a~fraction
of their neighbors [\citet{Buhl94}; \citet{Thompson88}; \citet{Reyes98}; \citet{Feldmeyer99}; \citet{Gupta00};
\citet{FeldmeyerSakmann00}; \citet{PetersenSakmann00}; \citet{Binzegger04}; \citet{Song2005}; \citet{Mishchenko2009b}],
implying that many elements of the connectivity matrix $\mathbf{w}$
are zero;
see also [\citet{PAN04c}; \citet{Rigat06}; \citet{PILL07}; \citet{Stevenson08}] for further
discussion. Second, ``Dale's law'' states that each of a neuron's
postsynaptic connections in the adult cortex (and many other brain
areas) must all be of the same sign (either excitatory or inhibitory).
Both of these priors are easy to incorporate in the M-step
optimization, as we discuss below.

\subsubsection{Imposing a sparse prior on the connectivity}

It is well known that imposing sparseness via an $\mathrm{L}_1$-regularizer can
dramatically reduce the amount of data necessary to accurately
reconstruct sparse high-dimensional parameters
[\citet{Tibs96}; \citet{TIP01}; \citet{DE03}; \citet{NG04}; \citet{Candes2008}; \citet{Mishchenko2009}]. We
incorporate a prior of the form $\ln p(\mathbf{w}) = \mathit{const} - \lambda
\sum_{i,j} |w_{ij}|$, and additionally enforce the constraints~\mbox{$|w_{ij}|<L$}, for a suitable constant $L$ (since both excitatory and
inhibitory cortical connections are known to be bounded in
size). Since the penalty $\ln p(\mathbf{w})$ is concave, and the constraints
$|w_{ij}|<L$ are convex, we may solve the resulting optimization
problem in the M-step using standard convex optimization methods
[\citet{CONV04}]. In addition, the problem retains its separable
structure: the full optimization may be broken up into $N$ smaller
problems that may be solved independently.

\subsubsection{Imposing Dale's law on the connectivity}

Enforcing Dale's law requires us to solve a nonconvex, nonseparable
problem: we need to optimize the concave function $Q(\mathbf{\theta
},\mathbf{\theta}^{(l)})
+ \ln P(\mathbf{\theta})$ under the nonconvex, nonseparable
constraint that all
of the elements in any column of the matrix $\mathbf{w}$ are of the
same sign
(either nonpositive or nonnegative). It is difficult to solve this
nonconvex problem exactly, but we have found that simple greedy
methods are quite efficient in finding good approximate solutions.

We begin with our original sparse solution, obtained as discussed in
the previous subsection without enforcing Dale's law. Then we assign
each neuron as either excitatory or inhibitory, based on the weights
we have inferred in the previous step: that is, neurons $i$ whose
inferred postsynaptic connections $w_{ij}$ are largely positive are
tentatively labeled excitatory, and neurons with largely inhibitory
inferred postsynapic connections are labeled inhibitory. Neurons which
are highly ambiguous may be unassigned in the early iterations, to
avoid making mistakes from which it might be difficult to
recover. Given the assignments $a_i$ ($a_i =1$ for putative excitatory
cells, $-1$ for inhibitory, and $0$ for neurons which have not yet
been assigned), we solve the convex, separable problem
%
%
\begin{equation}
\mathop{\operatorname{argmax}}_{a_i w_{ij} \geq0, |w_{ij}|<L
\forall i,j}
Q\bigl(\mathbf{\theta},\mathbf{\theta}^{(l)}\bigr) - \lambda\sum_{ij} |w_{ij}|,
\end{equation}
which
may be handled using the standard convex methods discussed
above. Given the new estimated connectivities $\mathbf{w}$, we can re-assign
the labels $a_i$, or flip some randomly to check for local optima. We
have found this simple approach to be effective in practice.

\subsection{Specific implementation notes} \label
{sec:methods:specific_implementation}

Pseudocode summarizing our approach is given in Algorithm \ref
{eqn:pseudocode}. As discussed in Section \ref{sec:methods:indep}, the
intrinsic parameters $\tilde{\mathbf{\theta}}_i$ may be initialized
effectively using the
methods described in [\citet{Vogelstein2009}]; then the full parameter
$\mathbf{\theta}$ is estimated via EM, where we use the
embedded-chain-within-blockwise-Gibbs approach discussed in Section
\ref
{sec:methods:joint} (or the cheaper factorized approximation described
in Section~\ref{sec:cheaper-high-snr}) to obtain the sufficient
statistics in the E-step and the separable convex optimization methods
discussed in Section \ref{sec:methods:parametersHMM} for the M-step.


\begin{algorithm}
\caption{Pseudocode for estimating connectivity from
calcium imaging data using EM; $\eta_1$ and $\eta_2$ are
user-defined convergence tolerance parameters.}
\label{eqn:pseudocode}
\hspace*{15pt}\textbf{while} $|{\mathbf{w}}^{(l)}-{\mathbf{w}}^{(l-1)}|>\eta_1$ \textbf{do}\\
\hspace*{30pt}\textbf{for all} $i=1\ldots N$ \textbf{do}\\
\hspace*{45pt}\textbf{while} $|{\tilde{\mathbf{\theta}}_i}^{(l)}-{\tilde{\mathbf{\theta
}}_i}^{(l-1)}|> \eta_2$ \textbf{do}\\
\hspace*{60pt}Approximate $P[X_i(t)|F_i; \tilde{\mathbf{\theta}}_i]$ using
SMC (Section \protect\ref{sec:methods:indep})\\
\hspace*{60pt}Perform the M-step for the intrinsic parameters $\tilde{\mathbf
{\theta}}_i$ (Section \protect\ref{sec:methods:indep})\\
\hspace*{45pt}\textbf{end while}\\
\hspace*{30pt}\textbf{end for}\\
\hspace*{30pt}\textbf{for all} $i=1\ldots N$ \textbf{do}\\
\hspace*{45pt}Approximate $P[n_i(t), \mathbf{h}_i(t) |\mathbf{F}; \mathbf
{\theta}_i]$ using either the
blockwise Gibbs\\
\hspace*{45pt}method or the factorized approximation (Section
 \ref{sec:methods:joint})\\
\hspace*{30pt}\textbf{end for}\\
\hspace*{30pt}\textbf{for all} $i=1\ldots N$ \textbf{do}\\
\hspace*{45pt}Perform the M-step for $\{b_i, \mathbf{w}_i\}_{i\leq N}$ using
separable convex optimization\\
\hspace*{45pt}methods (Section \protect\ref
{sec:methods:parametersHMM})\\
\hspace*{30pt}\textbf{end for}\\
\hspace*{15pt}\textbf{end while}
\end{algorithm}

As emphasized above, the parallel nature of these EM steps is
essential for making these computations tractable. We performed the
bulk of our analysis on a 256-processor cluster of Intel Xeon L5430
based computers (2.66~GHz). For 10~minutes of simulated fluorescence
data, imaged at $30$~Hz, calculations using the factorized
approximation typically took 10--20~mi\-nutes per neuron (divided by the
number of available processing nodes on the cluster), with time split
approximately equally between (i) estimating the intrinsic parameters
$\tilde{\mathbf{\theta}}_i $, (ii) approximating the posteriors
using the independent
SMC method, and (iii) estimating the connectivity matrix,
$\mathbf{w}$. The hybrid embedded-chain-within-blockwise-Gibbs sampler was
substantially slower, up to an hour per neuron, with the Gibbs sampler
dominating the computation time, because we thinned the chain by a
factor of five, following preliminary quantification of the
autocorrelation timescale of the Gibbs chain (data not shown).

\subsection{Simulating a neural population} \label{sec:results:simulations}

To test the described method for inferring connectivity from calcium
imaging data, we simulated networks of spontaneously firing randomly
connected neurons according to our model,
equations~(\ref{eqn:glm:definition})--(\ref{eqn:F:definition}), and also
using other network models (see Section \ref{sec:impact-strong-corr}).
Although simulations ran at $1$ msec time discretization, the imaging
rate was assumed to be much slower: $5$--$200$ Hz
(cf. Figure \ref{fig:recvar} below).

Model parameters were chosen based on experimental data available in
the literature for cortical neural networks
[\citet{Sayer1990}; \citet{Braitenberg1998}; \citet{Urquijo2000}; \citet{Lefort2009}]. More
specifically, the network consisted of 80\% excitatory and 20\%
inhibitory neurons [\citet{Braitenberg1998}; \citet{Urquijo2000}], each respecting
Dale's law (as discussed in Section \ref{sec:methods:parametersHMM}
above). Neurons were randomly connected to each other in a spatially
homogeneous manner with probability $0.1$
[\citet{Braitenberg1998}; \citet{Lefort2009}]. Synaptic weights for excitatory
connections, as defined by excitatory postsynaptic potential (PSP)
peak amplitude, were randomly drawn from an exponential distribution
with the mean of $0.5$ mV [\citet{Lefort2009}; \citet{Sayer1990}]. Inhibitory
connections were also drawn from an exponential distribution, their
strengths chosen so as to balance excitatory and inhibitory currents
in the network, and achieve an average firing rate of $\approx$5 Hz
[\citet{Abeles91}]. Practically, this meant that the mean strength of
inhibitory connections was about 10 times larger than that of the
excitatory connections. PSP shapes were modeled as an alpha function
[\citet{Koch99}]: roughly, the difference of two exponentials,
corresponding to a~sharp rise and relatively slow decay
[\citet{Sayer1990}]. We neglected conduction delays, given that the time
delays below $\sim$1 msec expected in the local cortical circuit were
far below the time resolution of our simulated imaging data.

Note that PSP peak amplitudes measured {in vitro} [as in, e.g.,
\citet{Song2005}] cannot be incorporated directly in
equation \eqref{eqn:glm:definition}, since the synaptic weights in our
model---$w_{ij}$ in equation \eqref{eqn:glm:definition}---are
dimensionless quantities representing the change in the spiking
probability of neuron $i$ given a spike in neuron $j$, whereas PSP
peak amplitude describes the physiologically measured change in the
membrane voltage of a neuron due to synaptic currents triggered by a
spike in neuron $j$. To relate the two, note that in order to trigger
an immediate spike in a neuron that typically has its membrane voltage
$V_b$ mV below the spiking threshold, roughly $n_E = V_b / V_E$
simultaneous excitatory PSPs with the peak amplitude~$V_E$ would be
necessary. Therefore, the change in the spiking probability of a
neuron due to excitatory synaptic current~$V_E$ can be approximately
defined as
%
%
\begin{equation}\label{eqn:convert:leadin-1}
\delta P_E = V_E/V_b
\end{equation}
(so that $\delta P_E n_E \approx1$). $V_b \approx15$ mV here, while
values for the PSP amplitude $V_E$ were chosen as described above.
Similarly, according to equation (\ref{eqn:glm:definition}), the same change
in the spiking probability of a neuron $i$ following the spike of a
neuron~$j$ in the GLM is roughly
%
%
\begin{equation}\label{eqn:convert:leadin-2}
\delta P_E = [ f(b_i + w_{ij}) - f(b_i) ] \tau_h,
\end{equation}
where recall $\tau_h$ is the typical PSP time-scale, that is, the time
over which a~spike in neuron $j$ significantly affects the firing
probability of the neuron~$i$. Equating these two expressions gives
us a simple method for converting the physiological parameters $V_E$
and $V_b$ into suitable GLM parameters $w_{ij}$.

Finally, parameters for the internal calcium dynamics and fluorescence
observations were chosen according to our experience with several
cells analyzed using the algorithm of [\citet{Vogelstein2009}], and
conformed to previously published results
[\citet{ImagingManual}; \citet{HelmchenSakmann96}; \citet{BrenowitzRegehr07}]. Table
\ref{table:caparm} summarizes the details for each of the parameters
in our model.

\begin{table}[t]
\tablewidth=\textwidth
\caption{Table of simulation parameters} \label{table:caparm}
\begin{tabular*}{\textwidth}{@{\extracolsep{\fill}}lcc@{}}
\hline
\textbf{Variable} & \textbf{Value/distribution} & \textbf{Unit} \\
\hline
Total neurons & 10--500 & \# \\
Excitatory neurons & $80$ & $\%$ \\
Connections sparseness & $10$ & $\%$ \\
Baseline firing rate & $5$ & Hz\\[3pt]
Excitatory PSP peak height & $\sim\mathcal{E}(0.5)$ & mV \\
Inhibitory PSP peak height & $\sim-\mathcal{E}(2.3)$ & mV \\
Excitatory PSP rise time & 1 & msec \\
Inhibitory PSP rise time & 1 & msec \\
Excitatory PSP decay time & $\sim\mathcal{N}_{0.5}(10,2.5)$ & msec \\
Inhibitory PSP decay time & $\sim\mathcal{N}_{0.5}(20,5)$ & msec\\
Refractory time, $w_{ii}$ & $\sim\mathcal{N}_{0.5}(10,2.5)$ & msec \\[3pt]
Calcium std. $\sigma_c$ & $\sim\mathcal{N}_{0.4}(28,10)$ & \textmu M\\
Calcium jump after spike, $A_c$ & $\sim\mathcal{N}_{0.4}(80,20)$ &
\textmu M\\
Calcium baseline, $C_b$ & $\sim\mathcal{N}_{0.4}(24,8)$ & \textmu M\\
Calcium decay time, $\tau_c$ & $\sim\mathcal{N}_{0.4}(200,60)$ &
msec\\
Dissociation constant, $K_d$ & $200$ & \textmu M \\[3pt]
Fluorescence scale, $\alpha$ & $1$ & n/a\\
Fluorescence baseline, $\beta$ & $0$ & n/a\\
Signal-dependent noise, $\gamma$ & $10^{-3}-10^{-5}$ & n/a\\
Signal-independent noise, $\sigma^F$ & $4\cdot10^{-3}-4\cdot
10^{-5}$ & n/a\\
\hline
\end{tabular*}
\tabnotetext[]{}{$\mathcal{E}(\lambda)$
indicates an exponential distribution with mean $\lambda$, and
$\mathcal{N}_p(\mu,\sigma^2)$ indicates a normal distribution with
mean $\mu$ and variance $\sigma^2$, truncated at lower bound $p\mu$.
Units (when applicable) are given with respect to mean values (i.e.,
units are squared for variance).}
\end{table}

\section{Results}
\label{sec:results}

\begin{figure}

\includegraphics{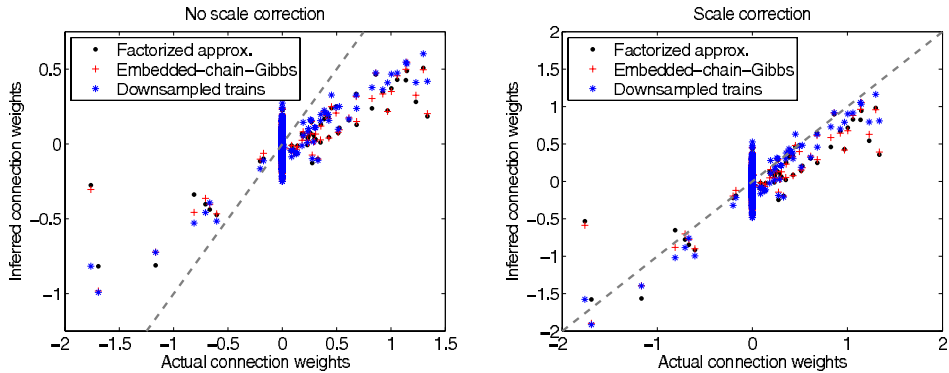}

\caption{Quality of the connectivity matrix estimated from simulated
calcium imaging data. Inferred connection weights $\hat w_{ij}$ are
shown in a scatter plot versus real connection weights $w_{ij}$, with
inference performed using the factorized approximation, exact
embedded-chain-within-blockwise-Gibbs approach, and true spike trains
down-sampled to the frame rate of the calcium imaging. A network of
$N=25$ neurons was used, firing at $\approx$5~Hz, and imaged for
$T=10$ min at $60$ Hz with intermediate eSNR $\approx$ 6 [see
equation \textup{(\protect\ref{eq:eSNR})} and Figure \textup{\protect\ref{fig:recvar-SNR}} below]. The
squared correlation coefficient between the connection weights
calculated using the factorized approximation and true connection
weights was $r^2=0.47$, compared with the
embedded-chain-within-blockwise-Gibbs method's $r^2=0.48$. For
connection weights calculated directly from the true spike train
down-sampled to the calcium imaging frame rate, we obtained $r^2=0.57$.
(For comparison, $r^2=0.71$ for the connectivity matrix calculated
using the full spike trains with 1 ms precision; data not shown.)
Here and in the following figures the gray dashed line indicates
unity, $y=x$. The inferred connectivity in the left panel shows a
clear scale bias, which can be corrected by dividing by the scale
correction factor calculated in Section \textup{\protect\ref{sec:scale}} below (right
panel). The vertical lines apparent at zero in both subplots are due
to the fact that the connection probability in the true network was
significantly less than one: that is, many of the true weights $w_{ij}$
are exactly zero.}
\label{fig:scatters}
\end{figure}

In this section we study the performance of our proposed network
estimation methods, using the simulated data described in Section~%
\ref{sec:results:simulations} above. Specifically, we estimated the
connectivity matrix using both the
embedded-chain-within-blockwise-Gibbs approach and the simpler
factorized approximation. \mbox{Figure}~\ref{fig:scatters} summarizes one
typical experiment: the EM algorithm using the factorized
approximation estimated the connectivity matrix about as accurately as
the full embedded-chain-within-blockwise-Gibbs approach ($r^2=0.47$
vs. $r^2=0.48$). Thus, in the following we will focus primarily on
the factorized approximation, since this is much faster than the full
blockwise-Gibbs approach (recall Section~\ref{sec:methods:specific_implementation}).

%

\subsection{Impact of coarse time discretization of calcium imaging
data and scale factor of inferred connection weights}
\label{sec:scale}

A notable feature of the results illustrated in the left panel of
Figure \ref{fig:scatters} is that our estimator is biased downward by a
roughly constant scale factor: our estimates $\hat w_{ij}$ are
approximately linearly related to the true values of $w_{ij}$ in the
simulated network, but the slope of this linear relationship is less
than one. At first blush, this bias does not seem like a major
problem: as we discussed in Section \ref{sec:results:simulations},
even in the noiseless case we should at best expect our estimated
coupling weights $\hat w_{ij}$ to correspond to some monotonically
increasing function of the true neural connectivities, as measured by
biophysical quantities such as the peak PSP amplitude. Nonetheless,
we would like to understand the source of this bias more
quantitatively; in this section we discuss this issue in more depth
and derive a simple method for correcting the bias.

The bias is largely due to the fact that we suffer a loss of temporal
resolution when we attempt to infer spike times from slowly-sampled
fluorescence data. As discussed in [\citet{Vogelstein2009}], we can
recover some of this temporal information by using a finer time
resolution for our recovered spike trains than~$\Delta$, the time
resolution of the observed fluorescence signal. However, when we
attempted to infer $\mathbf{w}$ directly spike trains sampled from the
posterior $P[\mathbf{X}|\mathbf{F}]$ at \mbox{higher-than-$\Delta$}
resolution, we found
that the inferred connectivity matrix was strongly biased toward the
symmetrized matrix $(\mathbf{w}+\mathbf{w}^\mathrm{T})/2$ (data not shown). In
other words,
whenever a nearly synchronous jump was consistently observed in two
fluorescent traces $F_i(t)$ and $F_j(t)$ (at the reduced time
resolution~$\Delta$), the EM algorithm would typically infer an
excitatory \textit{bidirectional} connection: that is, both $\hat w_{ij}$
and $\hat w_{ji}$ would be large, even if only a unidirectional
connection existed between neurons $i$ and $j$ in the true network.
While we expect, by standard arguments, that the Monte Carlo EM
estimator constructed here should be consistent (i.e., we should
recover the correct $\mathbf{w}$ in the limit of large data length $T$ and
many Monte Carlo samples), we found that this bias persisted given
experimentally-reasonable lengths of data and computation time.

Therefore, to circumvent this problem, we simply used the original
imaging time resolution $\Delta$ for the inferred spike trains: note
that, due to the definition of the spike history terms $h_{ij}$ in
equation (\ref{eqn:h:definition}), a spike in neuron~$j$ at time $t$ will
only affect neuron $i$'s firing rate at time~$t+\Delta$ and greater.
This successfully counteracted the symmetrization problem (and also
sped the calculations substantially), but resulted in the scale bias
exhibited in Figure \ref{fig:scatters}, since any spikes that fall
into the same time bin are treated as coincidental: only spikes that
precede spikes in a neighboring neuron by at least one time step will
directly affect the estimates of~$w_{ij}$, and therefore grouping
asynchronous spikes within a single time bin $\Delta$ results in a
loss of information.

\begin{figure}

\includegraphics{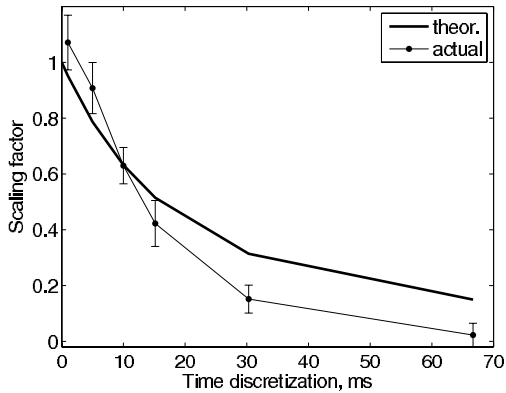}

\caption{The low frame rate of calcium imaging explains the scale
error observed in the inferred connectivity weights shown in Figure
\textup{\protect\ref{fig:scatters}}. A correction scale factor may be calculated
analytically (thick line) as discussed in the main text
[equation (\protect\ref{eqn:bias})]. The scale error observed empirically (thin
line) matches well with this theoretical estimate. In the latter case,
the scale error was calculated from the fits obtained directly from
the true spike trains, down sampled to different $\Delta$, for a
network of $N=25$ neurons firing at $\approx$5 Hz and observed for
$T=10$ min. The error-bars indicate 95\% confidence intervals for
scale error at each $\Delta$.}
\label{fig:bias}
\end{figure}

To estimate the magnitude of this time-discretization bias more
quantitatively, we consider a significantly simplified case of two
neurons coupled with a small weight $w_{12}$, and firing with baseline
firing rate of $r=f(b)$. In this case an approximate sufficient
statistic for estimating $w_{12}$ may be defined as the expected
elevation in the spike rate of neuron one on an interval of length
$\mathcal{T}$, following a spike in neuron two:
%
%
\begin{eqnarray}\label{eqn:scale:leadin-1}
SS   &=& E\biggl[\int _{t'}^{t'+\mathcal{T}} n_1(t)\, dt
\big|n_2(t')=1, n_2(t)=0 \mbox{ }  \forall
t \in(t',t'+\mathcal{T}] \biggr]
\nonumber
\\[-8pt]
\\[-8pt]
\nonumber
&\approx& r
\mathcal{T} + f'(b) w_{12}\tau_h,
\end{eqnarray}
where $f'(b)$ represents the slope of the nonlinear function $f(\cdot)$ at
the baseline level $b$. This approximation leads to a conceptually
simple method-of-moments estimator,
%
%
\begin{equation}\label{eqn:scale:leadin-2}
\hat w_{12}=(SS-r\mathcal{T})/f'(b)\tau_h.
\end{equation}
Now, if the spike trains are down-sampled into time-bins of size
$\Delta$,
we must estimate the statistic $SS$ with a discrete sum instead:
%
%
\begin{eqnarray}\label{eqn:scale:leadin-3}
SS^{ds}  & = &E\Biggl[\sum _{t=t'+\Delta}^{t'+\Delta+
\mathcal{T}} n^{\mathit{ds}}_1(t)   \Big|   n^{\mathit{ds}}_2(t')=1, n^{\mathit{ds}}_2(t)=0
\mbox{ }\forall t \in(t',t'+\mathcal{T}]\Biggr] \nonumber\\
& \approx& r \mathcal{T}
+ f'(b) \int _0^\Delta\frac{dt'}{\Delta}
\int _{\Delta}^{\Delta+ \mathcal{T}}
w_{12}\exp\bigl(-(t-t')/\tau_h\bigr)\, dt \\
&\approx& r \mathcal{T} +
f'(b)w_{12}\frac{1-\exp(-\Delta/\tau_h)}{\Delta/\tau_h^2}.\nonumber
\end{eqnarray}
$n^{\mathit{ds}}(t)$ here are down-sampled spikes, that is, the spikes defined
on a
grid $t=0,\Delta,2\Delta,\ldots.$ In the second equality we made the
approximation that the true position of the spike of the second
neuron, $n^{\mathit{ds}}_2(t')$, may be uniformly distributed in the first
time-bin $[0,\Delta]$, and the discrete sum over $t$ is from the
second time-bin $[\Delta,2\Delta]$ to
$[\mathcal{T},\mathcal{T}+\Delta]$, that is, over all spikes of the first
neuron that occurred in any of the strictly subsequent time-bins up to
$\mathcal{T} + \Delta$. Forming a method-of-moments estimator as in
equation (\ref{eqn:scale:leadin-2}) leads to a biased estimate,
%
%
\begin{equation}\label{eqn:bias}
\hat w_{12}^{\mathit{ds}}\approx\frac{1-\exp(-\Delta/\tau_h)}{\Delta/\tau_h}
\hat w_{12},
\end{equation}
and somewhat surprisingly (given the rather crude nature of these
approximations), this corresponds quite well with the scale bias we
observe in practice. In Figure~\ref{fig:bias} we plot the scale bias
from equation (\ref{eqn:bias}) versus that empirically deduced from our
simulations for different values of $\Delta$; we see that
equation (\ref{eqn:bias}) describes the observed scale bias fairly well.
Thus, we can divide by this analytically-derived factor to effectively
correct the bias of our estimates, as shown in the right panel of
Figure \ref{fig:scatters}.

\subsection{Impact of prior information on the inference}

\begin{figure}

\includegraphics{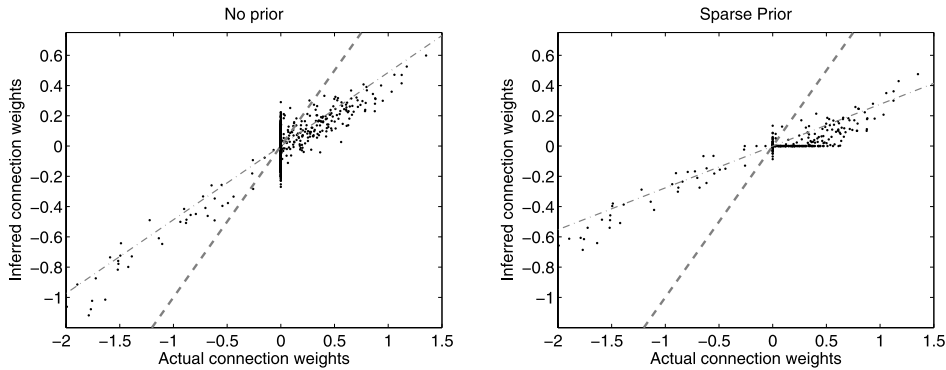}

\caption{Imposing a sparse prior on connectivity improves our
estimates. Scatter plots indicate the connection weights $w_{ij}$
reconstructed using no prior ($r^2=0.64$; left panel) and a sparse
prior ($r^2=0.85$; right panel) vs. the true connection weights in
each case. These plots were based on a simulation of $N=50$ neurons
firing at $\approx$5 Hz, imaged for $T=10$ min at $60$ Hz, with eSNR
$\approx10$. Clearly, the sparse prior reduces the relative error,
as indicated by comparing the relative distance between the data
points (black dots) to the best linear fit (gray dash-dotted line), at
the expense of some additional soft-threshold bias, as is usual in the
L$_1$ setting.}
\label{fig:sparse}
\end{figure}

\begin{figure}

\includegraphics{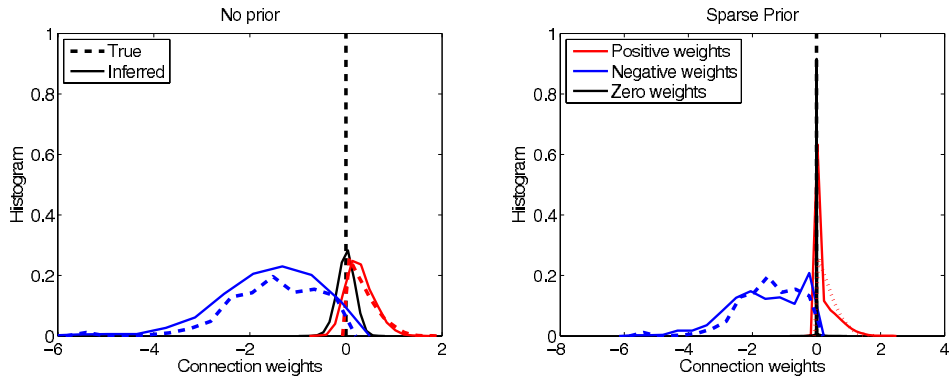}

\caption{The distributions of inferred connection weights using no
prior (left panel) and a sparse prior (right panel) vs. true
distributions. When the sparse prior is enforced, zero weights are
recovered with substantially higher frequency (black lines), thus
allowing better identification of connected neural pairs. Likewise,
excitatory and inhibitory weights are better recognized (red and blue
lines, respectively), thus allowing accurate classification of neurons
as excitatory or inhibitory. The normalized Hamming distance between
the inferred and true connectivity matrix here (defined as $H(\mathbf{w},
\hat\mathbf{w}) = [N (N-1)]^{-1} \sum_{ij} |\operatorname{sign} (w_{ij}) -
\operatorname{sign} (\hat w_{ij})|$, with the convention
$\operatorname{sign}(0)=0$)
was~$0.06$. Distributions are shown for a simulated population of
$N=200$ neurons firing at $\approx$5 Hz and imaged for $T=10$ min at
$60$ Hz, with eSNR $\approx$10. Note that the peak at zero in the
true distributions (black dashed trace) corresponds to the vertical
line visible at zero in Figures \textup{\protect\ref{fig:scatters}} and
\textup{\protect\ref{fig:sparse}}; inferred distributions were rescaled to remove bias, as
described in Section~\textup{\protect\ref{sec:scale}}.}
\label{fig:distros}
\end{figure}

Next we investigated the importance of incorporating prior information
in our estimates. We found that imposing a sparse prior (as described
in Section \ref{sec:methods:parametersHMM}) significantly improved
our results. For example, Figure \ref{fig:sparse} illustrates a case in
which our obtained $r^2$ increased from $0.64$ (with no L$_1$
penalization in the M-step) to $0.85$ (with penalization; the penalty
$\lambda$ was chosen approximately as the inverse mean absolute value
of $w_{ij}$, which is known here because we prepared the network
simulations, but is available in practice given the previous
physiological measurements discussed in Section
\ref{sec:results:simulations}). See also Figure \ref{fig:recvar-NT}
below. Furthermore, the weights estimated using the sparse prior more
reliably provide the sign (i.e., excitatory or inhibitory) of each
presynaptic neuron in the network (Figure \ref{fig:distros}).

Incorporation of Dale's law, on the other hand, only leads to an
$\approx$10\% change in the estimation $r^2$ in the absence of an
L$_1$ penalty, and no significant improvement at all in the presence
of an L$_1$ penalty (data not shown). Thus, Dale's prior was not
pursued further here.
%

\subsection{Impact of experimental factors on estimator
accuracy}\label
{sec:results-parm}

\begin{figure}

\includegraphics{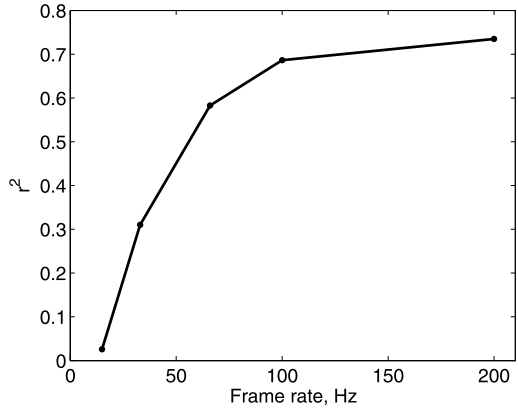}

\caption{Accuracy of the inferred connectivity as a function of the
frame rate of calcium imaging. A~population of $N=25$ neurons firing
at $\approx$5 Hz and imaged for $T=10$ min was simulated here, with
eSNR $\approx10$. At $100$ Hz, $r^2$ saturated at the level
$r^2\approx0.7$ achieved with $\Delta\rightarrow0$.}
\label{fig:recvar}
\end{figure}

\begin{figure}

\includegraphics{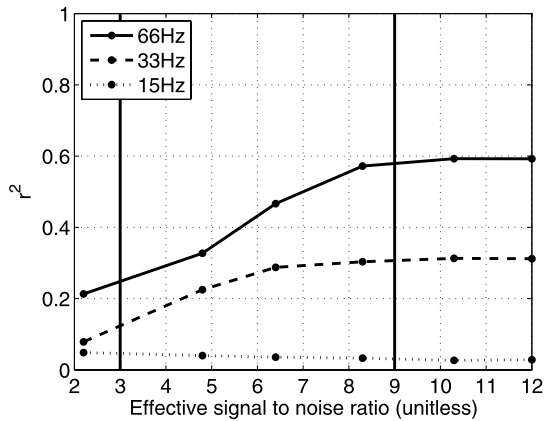}

\caption{Accuracy of inferred connectivity as a function of effective
imaging SNR [eSNR, defined in equation \textup{(\protect\ref{eq:eSNR})}], for frame rates of
\textup{15, 33} and \textup{66} Hz. Neural population simulation was the same as in
Figure \textup{\protect\ref{fig:recvar}}. Vertical black lines correspond to the eSNR
values of the two example traces in Figure \textup{\protect\ref{fig:example_traces}},
for comparison.}
\label{fig:recvar-SNR}
\end{figure}

Next we\break sought to quantify the minimal experimental conditions
necessary for accurate estimation of the connectivity matrix. Figure
\ref{fig:recvar} shows the quality of the inferred connectivity matrix
as a function of the imaging frame rate, and indicates that imaging
frame rates~$\geq$ 30 Hz are needed to achieve meaningful
reconstruction results. This matches nicely with currently-available
technology; as discussed in the introduction, $30$ or~$60$~Hz imaging
is already in progress in a number of laboratories
[\citet{NguyenParker01}; \citet{Iyer06}; \citet{SalomeBourdieu06}; \citet{ReddySaggau08}], though in
some cases higher imaging rates come at a cost in the signal-to-noise
ratio of the images or in the number of neurons that may be imaged
simultaneously. Similarly, Figure \ref{fig:recvar-SNR} illustrates
the quality of the inferred connectivity matrix as a function of the
effective SNR measure defined in equation~(\ref{eq:eSNR}).

\begin{figure}

\includegraphics{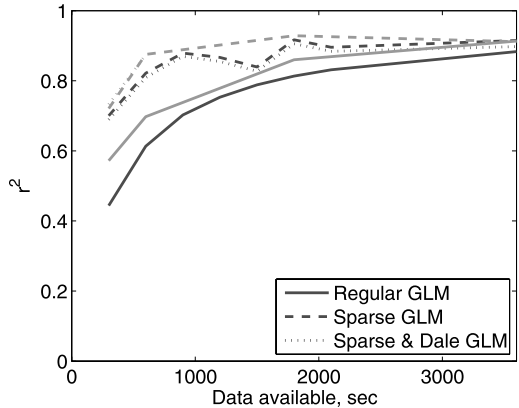}

\caption{Accuracy of inferred connectivity as a function of the
imaging time and neural population size. Incorporating a sparse prior
dramatically increases the reconstruction quality (dashed lines). When
the sparse prior is imposed, $T=5$ min is sufficient to recover $70\%$
of the variance in the connection weights. Incorporating Dale's prior
leads to only marginal improvement (dotted line). Furthermore,
reconstruction accuracy does not strongly depend on the neural
population size, $N$. Here, neural populations of size $N=100$ and
$200$ are shown (black and gray, respectively), with eSNR $\approx10$
and $60$ Hz imaging rate in each case.}
\label{fig:recvar-NT}
\vspace*{-6pt}
\end{figure}

Finally, Figure \ref{fig:recvar-NT} shows the quality of the inferred
connectivity matrix as a function of the experimental duration. The
minimal amount of data for a particular~$r^2$ depended substantially
on whether the sparse prior was enforced. In particular, when not
imposing a sparse prior, the calcium imaging duration necessary to
achieve $r^2=0.5$ for the reconstructed connectivity matrix in this
setting was $T\approx10$ min, and $r^2=0.75$ was achieved at
$T\approx30$~min. With a sparse prior, $r^2>0.7$ was achieved
already at $T\approx5$~min. Furthermore, we observed that the
accuracy of the reconstruction did not deteriorate dramatically with
the size of the imaged neural population: roughly the same
reconstruction quality was observed (given a fixed length of data) for
$N$ varying between $50$--$200$ neurons. These results were
consistent with a~rough Fisher information computation which we
performed but have omitted here to conserve space.

\subsection{Impact of strong correlations and deviations from
generative model on the inference}
\label{sec:impact-strong-corr}

\begin{figure}

\includegraphics{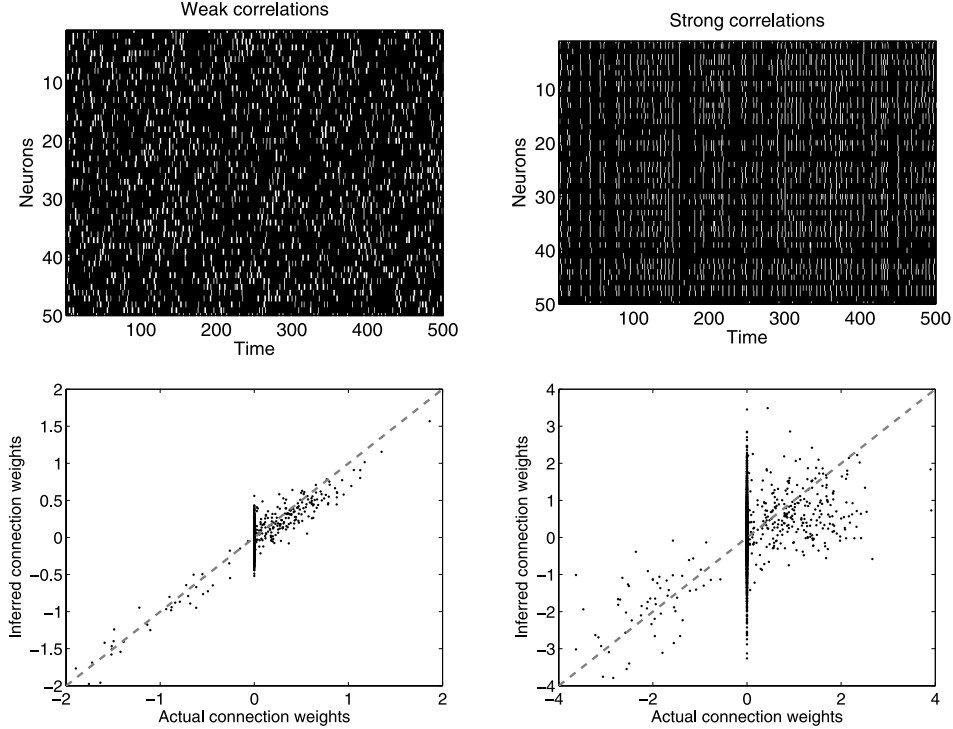}

\caption{Diversity of observed neural activity patterns is required
for accurate circuit inference. Here, 15 sec of simulated spike trains
for a weakly coupled network (top left panel) and a network with
strongly coupled component (top right panel) are shown. In weakly
coupled networks, spikes are sufficiently uncorrelated to give access
to enough different neural activity patterns to estimate the weights
$\mathbf{w}$. In a strongly coupled case, many highly synchronous events
are evident (top right panel), thus preventing observation of a
sufficiently rich ensemble of activity patterns. Accordingly, the
connectivity estimates for the strongly coupled neural network (bottom
right panel) do not represent the true connectivity of the circuit,
even for the weakly coupled component. This is contrary to the
weakly-coupled network (bottom left panel) where true connectivity is
successfully obtained. Networks of $N=50$ neurons firing at $\approx$5
Hz and imaged for $T=10$ min at $60$ Hz were used to produce this
figure; eSNR $\approx10$.}
\label{fig:rasters}
\end{figure}

Estimation of network connectivity is fundamentally rooted in
observing changes in the spike rate conditioned on the state of the
other neurons. Considered from the point of view of estimating a
standard GLM, it is clear that the inputs to our model
(\ref{eqn:glm:definition}) must satisfy certain basic identifiability
conditions if we are to have any hope of accurately estimating the
parameter $\mathbf{w}$. In particular, we must rule out highly
multicollinear inputs $\{h_{ij}(t)\}$: speaking roughly, the set of
observed spike trains should be rich enough to span all $N$ dimensions
of $\mathbf{w}_i$, for each cell $i$. In the simulations pursued here,
the coupling matrix $\{w_{ij}\}_{i \neq j}$ was fairly weak and
neurons fired largely independently of each other: see
Figure \ref{fig:rasters}, upper left, for an illustration. In this case
of weakly-correlated firing, the inputs $\{h_{ij}(t)\}$ will also be
weakly correlated, and the model should be identifiable, as indeed we
found. Should this weak-coupling condition be violated, however
(e.g., due to high correlations in the spiking of a few neurons), we
may require much more data to obtain accurate estimates due to
multicollinearity problems.

To explore this issue, we carried out a simulation of a hypothetical
strongly coupled neural network, where, in addition to the
physiologically-relevant weak sparse connectivity discussed in Section
\ref{sec:results:simulations}, we introduced a sparse random strong
connectivity component. More specifically, we allowed a fraction of
neurons to couple strongly to the other neurons, making these
``command'' neurons which in turn could strongly drive the activity of
the rest of the population [\citet{MACLEAN05}]. The strength of this
strong connectivity component was chosen to dynamically build up the
actual firing rate from the baseline rate of $f(b) \approx1$ Hz to
approximately $5$ Hz. Such a network showed patterns of activity very
different from the weakly coupled networks inspected above (Figure
\ref{fig:rasters}, top right). In particular, a large number of highly
correlated events across many neurons were evident in this network. As
expected, our algorithm was not able to identify the true connectivity
matrix correctly in this scenario (Figure \ref{fig:rasters}, bottom
right panel). For ease of comparison, the left panels show a
``typical'' network (i.e., one lacking many strongly coupled neurons),
and its associated connectivity inference.

\begin{figure}

\includegraphics{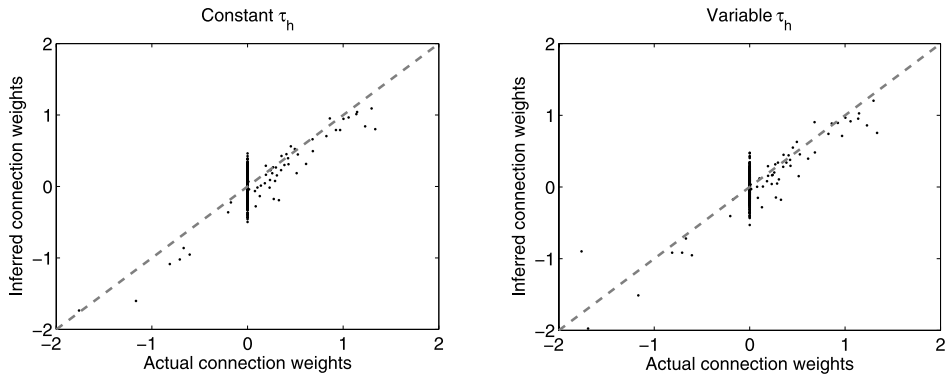}

\caption{Inference is robust to deviations of the data from our
generative model. With up to $25\%$ variability allowed in PSP time
scales $\tau_h$ (right panel), our algorithm provided reconstructions
of almost the same quality as when all $\tau_h$'s were the same (left
panel). Simulation conditions were the same as in Figure
\protect\ref{fig:recvar}, at $60$ Hz imaging rate.}
\label{fig:vartau}
\end{figure}

On the other hand, our inference algorithm showed significant
robustness to model misspecifcation, that is, deviations from our
generative model. One important such deviation is variation in the
time scales of PSPs in different synapses. Up to now, all PSP
time-scales were assumed to be the same, that is,
$\{\tau^h_{ij}\}_{i,j\leq N}=\tau_h$. In Figure \ref{fig:vartau} we
introduce additional variability in $\tau_h$ from one neuron to
another. Variability in $\tau_h$ results in added variance in the
estimates of the connectivity weights, $w_{ij}$, through the
$\tau_h$-dependence of the scaling factor equation \eqref{eqn:bias}.
However, we found that this additional variance was relatively
insignificant in cases where $\tau_h$ varied up to $25\%$ from neuron
to neuron. We also found that inference was robust to changes in the
sparseness of the underlying connectivity matrix: we simulated neural
populations of size $N=25$ and $N=50$ neurons, as above, with
connection sparseness varying from 5\% (very sparse) to 100\%
(all-to-all), and in all cases the performance of our algorithm
remained stable, with $r^2\approx0.9$ for the estimate of the
connected weights, $w_{ij}\neq0$ (data not shown). Finally,
simulations with more biophysically-based conductance-driven noisy
integrate-and-fire network models [\citet{Vogels05}] led to qualitatively
similar results, further establishing the robustness of these methods;
again, details are omitted to conserve space.

\section{Discussion}
\label{sec:discussion}
In this paper we develop a Bayesian approach for inferring
connectivity in a network of spiking neurons observed using calcium
fluorescent imaging. A number of previous authors have addressed the
problem of \mbox{inferring} neuronal connectivity given a fully-observed set
of spike trains in a network
[\citet{BRIL88}; \citet{CSK88}; \citet{BRIL92}; \citet{PAN03d}; \citet{PAN04c}; \citet{TRUC05}; \citet{Rigat06}; \citet{NYK06};
\citet{KP06}; \citet{Vidne08}; \citet{Stevenson2009}; \citet{Garofalo09}; \citet{Cocco09}],
but the main challenge in the present work is the indirect nature of
the calcium imaging data, which provides only noisy, low-pass
filtered, temporally sub-sampled observations of spikes of individual
neurons. To solve this problem, we develop a specialized
blockwise-Gibbs sampler that makes use of an embedded Markov chain
method due to [\citet{NBR03}]. The connectivity matrix is then inferred
in an EM framework; the M-step parallelizes quite efficiently and
allows for the easy incorporation of prior sparseness information,
which significantly reduces data requirements in this context. We have
found that these methods can effectively infer the connectivity in
simulated neuronal networks, given reasonable lengths of data,
computation time, and assumptions on the biophysical network
parameters.

To our knowledge, we are the first to address this problem using the
statistical deconvolution methods and EM formulation described here
[though see also \citet{Roxin08}, who fit simplified, low temporal
resolution transition-based models to the 10 Hz calcium data obtained
by \citet{IkegayaYuste04}]. However, we should note that
[\citet{Rigat06}]
developed a closely related approach to infer connectivity from
low-SNR electrical recordings involving possibly-misclassified spikes
(in contrast to the slow, lowpass-filtered calcium signals we discuss
here). In particular, these authors employed a very similar Bernoulli
GLM and developed a Metropolis-within-Gibbs sampler to approximate the
necessary sufficient statistics for their model. In addition
[\citet{Rigat06}] develop a more intricate hierarchical prior for the
connectivity parameter $\mathbf{w}$; while we found that a simple L$_1$
penalization was quite effective here, it will be worthwhile to
explore more informative priors in future work.

A number of possible improvements of our method are available. One of
the biggest challenges for inferring neural connectivity from
functional data is the presence of indirect inputs from unobserved
neurons [\citet{Nykamp05}; \citet{NYK06}; \citet{KP06}; \citet{Vidne08}; \citet{Vakorin09}]: it is typically
impossible to observe the activity of all neurons in a given circuit,
and correlations in the unobserved inputs can mimic connections among
different observed neurons. Developing methods to cope with such
unobserved common inputs is currently an area of active research, and
should certainly be incorporated in the methods we have developed
here.

Several other important directions for future work are worth noting.
First, recently-developed photo-stimulation methods for activating or
deactivating individual neurons or sub-populations
[\citet{Deisseroth05}; \citet{SzobotaIsacoff07}; \citet{Nikolenko08}] may be useful to
increase statistical power in cases where the circuit's unperturbed
activity may not allow reliable determination of a circuit's
connectivity matrix; in particular, by utilizing external stimulation,
we can in principle choose a sufficiently rich experimental design
(i.e., a sample of input activity patterns) to overcome the
multicollinearity problems discussed in the context of
Figure \ref{fig:rasters}.

Second, improvements of the algorithms for faster implementation are
under development. Specifically, fast nonnegative optimization-based
deconvolution methods may be a promising alternative
[\citet{Vogelstein08}; \citet{Pan08b}] to the SMC approach used here. In
addition, modifications of our generative model to incorporate
nonstationarities in the fluorescent signal (e.g., due to dye
bleaching and drift) are fairly straightforward.\looseness=-1

Third, a fully Bayesian algorithm for estimating the posterior
distributions of all the parameters (instead of just the MAP estimate)
would be of significant interest. Such a fully-Bayesian extension is
conceptually simple: we just need to extend our Gibbs sampler to
additionally sample from the parameter $\mathbf{\theta}$ given the
sampled spike
trains $\mathbf{X}$. Since we already have a method for drawing
$\mathbf{X}$ given
$\mathbf{\theta}$ and $\mathbf{F}$, with such an additional sampler
we may obtain
samples from $P(\mathbf{X},\mathbf{\theta}| \mathbf{F})$ simply by
sampling from $\mathbf{X}\sim
P(\mathbf{X}|\theta,\mathbf{F})$ and $\mathbf{\theta}\sim
P(\mathbf{\theta}|\mathbf{X})$, via blockwise-Gibbs.
Sampling from the posteriors $P(\mathbf{\theta}|\mathbf{X})$ in the
GLM setting is quite
tractable using hybrid Monte Carlo methods, since all of the necessary
posteriors are log-concave
[\cite{Ishwaran99}; \citet{Gamerman97}; \citet{Gamerman98}; \citet{Yashar08}].

Finally, most importantly, we are currently applying these algorithms
in preliminary experiments on real data. Checking the accuracy of our
estimates is of course more challenging in the context of
nonsimulated data, but a number of methods for partial validation are
available, including multiple-patch recordings [\citet{Song2005}],
photo stimulation techniques [\citet{Vovan07}], and fluorescent anatomical
markers which can distinguish between different cell types
[\citet{Meyer02}] (i.e., inhibitory vs.\ excitatory cells; cf.
Figure \ref{fig:distros}). We hope to present our results in the near
future.

\section*{Acknowledgments}
We thank R. Yuste, B. Watson, A. Packer, T. Sippy, T.~Mrsic-Flogel and V. Bonin for
data and helpful discussions, and A.~Rami\-rez for helpful comments on an earlier draft.


%

\printaddresses


\begin{thebibliography}{}

\bibitem[\protect\citeauthoryear{Abeles}{1991}]{Abeles91}
\textsc{Abeles, M.} (1991).
 \textit{Corticonics}.
  Cambridge Univ. Press, Cambridge.

\bibitem[\protect\citeauthoryear{Ahmadian, Pillow and Paninski}{2011}]{Yashar08}
\textsc{Ahmadian, Y., Pillow, J.} and \textsc{Paninski, L.} (2011).
  Efficient {Markov chain Monte Carlo} methods for decoding population
spike trains. \textit{Neural Comput.} \textbf{23}.

\bibitem[\protect\citeauthoryear{Andrieu, Doucet and Holenstein}{2007}]{Andrieu2007}
\textsc{Andrieu, C., Doucet, A.} and \textsc{Holenstein, A.} (2007).
 {Particle Markov chain Monte Carlo}. {Working paper}.

\bibitem[\protect\citeauthoryear{Bickel, Li and Bengtsson}{2008}]{BickelBengtsson08}
\textsc{Bickel, P., Li, B.} and \textsc{Bengtsson, T.} (2008).
  Sharp failure rates for the bootstrap particle filter in high
dimensions. In \textit{Pushing the
Limits of
Contemporary Statistics: Contributions in Honor of Jayanta K. Ghosh} (Clarke, B. and Ghosal, S., eds.) 318--329.
IMS, Beachwod, OH.
\MR{2459233}

\bibitem[\protect\citeauthoryear{Binzegger, Douglas and Martin}{2004}]{Binzegger04}
\textsc{Binzegger, T., Douglas, R.~J.} and \textsc{Martin, K.~A.~C.} (2004).
 {A quantitative map of the circuit of cat primary visual cortex}.
 \textit{J. Neurosci.} \textbf{24} 8441--8453.

\bibitem[\protect\citeauthoryear{Boyd and Vandenberghe}{2004}]{CONV04}
\textsc{Boyd, S.} and \textsc{Vandenberghe, L.} (2004).
 \textit{Convex Optimization}.
  Oxford Univ. Press.
\MR{2061575}

\bibitem[\protect\citeauthoryear{Boyden et~al.}{2005}]{Deisseroth05}
\textsc{Boyden, E.~S., Zhang, F., Bamberg, E., Nagel, G.} and \textsc{Deisseroth, K.} (2005).
  Millisecond-timescale, genetically targeted optical control
of neural
activity. \textit{Nat. Neurosci.} \textbf{8} 1263--1268.

\bibitem[\protect\citeauthoryear{Braitenberg and Schuz}{1998}]{Braitenberg1998}
\textsc{Braitenberg, V.} and \textsc{Schuz, A.} (1998).
 \textit{Cortex: Statistics and Geometry of Neuronal Connectivity.}
  Springer, Berlin.

\bibitem[\protect\citeauthoryear{Brenowitz and Regehr}{2007}]{BrenowitzRegehr07}
\textsc{Brenowitz, S.~D.} and \textsc{Regehr, W.~G.} (2007).
  Reliability and heterogeneity of calcium signaling at single
presynaptic boutons of cerebellar granule cells.
 \textit{J. Neurosci.} \textbf{27} 7888--7898.

\bibitem[\protect\citeauthoryear{Briggman and Denk}{2006}]{Briggman2006}
\textsc{Briggman, K.~L.} and \textsc{Denk, W.} (2006).
  Towards neural circuit reconstruction with volume electron microscopy
techniques.
 \textit{Curr. Opin. Neurobiol.} \textbf{16} 562.

\bibitem[\protect\citeauthoryear{Brillinger}{1988}]{BRIL88}
\textsc{Brillinger, D.} (1988).
  Maximum likelihood analysis of spike trains of interacting nerve
cells.
 \textit{Biol. Cybern.} \textbf{59} 189--200.

\bibitem[\protect\citeauthoryear{Brillinger}{1992}]{BRIL92}
\textsc{Brillinger, D.} (1992).
  Nerve cell spike train data analysis: A progression of technique.
 \textit{J. Amer. Statist. Assoc.} \textbf{87} 260--271.

\bibitem[\protect\citeauthoryear{Buhl, Halasy and Somogyi}{1994}]{Buhl94}
\textsc{Buhl, E., Halasy, K.} and \textsc{Somogyi, P.} (1994).
  Diverse sources of hippocampal unitary inhibitory postynaptic
potentials and the number of synaptic release sites.
 \textit{Nature} \textbf{368} 823--828.

\bibitem[\protect\citeauthoryear{Candes and Wakin}{2008}]{Candes2008}
\textsc{Candes, E.~J.} and \textsc{Wakin, M.} (2008).
  An introduction to compressive sampling.
 \textit{IEEE Signal Proc. Mag.} \textbf{25} 21--30.

\bibitem[\protect\citeauthoryear{Chornoboy, Schramm and Karr}{1988}]{CSK88}
\textsc{Chornoboy, E., Schramm, L.} and \textsc{Karr, A.} (1988).
  Maximum likelihood identification of neural point process systems.
 \textit{Biol. Cybern.} \textbf{59} 265--275.
\MR{0961117}

\bibitem[\protect\citeauthoryear{Cocco, Leibler and Monasson}{2009}]{Cocco09}
\textsc{Cocco, S., Leibler, S.} and \textsc{Monasson, R.} (2009).
 {Neuronal couplings between retinal ganglion cells inferred by
efficient inverse statistical physics methods}.
 \textit{Proc. Nat. Acad. Sci.}
 \textbf{106} 14058--14062.

\bibitem[\protect\citeauthoryear{Cossart, Aronov and Yuste}{2003}]{CAR03}
\textsc{Cossart, R., Aronov, D.} and \textsc{Yuste, R.} (2003).
  Attractor dynamics of network up states in the neocortex.
 \textit{Nature} \textbf{423} 283--288.

\bibitem[\protect\citeauthoryear{Dempster, Laird and Rubin}{1977}]{DLR77}
\textsc{Dempster, A., Laird, N.} and \textsc{Rubin, D.} (1977).
  Maximum likelihood from incomplete data via the {EM} algorithm.
 \textit{J. Roy. Statist. Soc. Ser. B} \textbf{39} 1--38.
\MR{0501537}

\bibitem[\protect\citeauthoryear{Djurisic et~al.}{2004}]{Djurisic04}
\textsc{Djurisic, M., Antic, S., Chen, W.~R.} and \textsc{Zecevic, D.} (2004).
  Voltage imaging from dendrites of mitral cells: {EPSP} attenuation
and spike trigger zones.
 \textit{J. Neurosci.} \textbf{24} 6703--6714.

\bibitem[\protect\citeauthoryear{Dombeck et~al.}{2007}]{DombeckTank07}
\textsc{Dombeck, D.~A., Khabbaz, A.~N., Collman, F., Adelman, T.~L.} and \textsc{Tank, D.~W.}
(2007).
  Imaging large-scale neural activity with cellular resolution in
awake, mobile mice.
 \textit{Neuron} \textbf{56} 43--57.

\bibitem[\protect\citeauthoryear{Donoho and Elad}{2003}]{DE03}
\textsc{Donoho, D.} and \textsc{Elad, M.} (2003).
  Optimally sparse representation in general (nonorthogonal)
dictionaries via {L}$^1$ minimization.
 \textit{PNAS} \textbf{100} 2197--2202.
\MR{1963681}

\bibitem[\protect\citeauthoryear{Douc, Cappe and Moulines}{2005}]{DCM05}
\textsc{Douc, R., Cappe, O.} and \textsc{Moulines, E.} (2005).
  Comparison of resampling schemes for particle filtering.
 In \textit{Proc. 4th Int. Symp. Image and Signal Processing and
 Analysis} 64--69. ISPA.

\bibitem[\protect\citeauthoryear{Doucet, de~Freitas and Gordon}{2001}]{DFG01}
\textsc{Doucet, A., de~Freitas, N.} and \textsc{Gordon, N.}, eds. (2001).
 \textit{Sequential Monte Carlo in Practice}.
  Springer, New York.

\bibitem[\protect\citeauthoryear{Doucet, Godsill and Andrieu}{2000}]{DGA00}
\textsc{Doucet, A., Godsill, S.} and \textsc{Andrieu, C.} (2000).
 {On sequential Monte Carlo sampling methods for Bayesian filtering}.
 \textit{Stat. Comput.} \textbf{10} 197--208.

\bibitem[\protect\citeauthoryear{Escola and Paninski}{2011}]{Escola07}
\textsc{Escola, S.} and \textsc{Paninski, L.} (2011).
  Hidden {M}arkov models applied toward the inference of neural states
and the improved estimation of linear receptive fields.
 \textit{Neural Comput.} To appear.

\bibitem[\protect\citeauthoryear{Feldmeyer et~al.}{1999}]{Feldmeyer99}
\textsc{Feldmeyer, D., Egger, V., Lubke, J.} and \textsc{Sakmann, B.} (1999).
  Reliable synaptic connections between pairs of excitatory
layer 4
neurones within a single ``barrel'' of developing rat somatosensory cortex.
 \textit{J. Physiol.} \textbf{1} 169--90.

\bibitem[\protect\citeauthoryear{Feldmeyer and Sakmann}{2000}]{FeldmeyerSakmann00}
\textsc{Feldmeyer, D.} and \textsc{Sakmann, B.} (2000).
  Synaptic efficacy and reliability of excitatory connections between
the principal neurones of the input (layer 4) and output (layer
5) of
the neocortex.
 \textit{J. Physiol.} \textbf{525} 31--39.

\bibitem[\protect\citeauthoryear{Gamerman}{1997}]{Gamerman97}
\textsc{Gamerman, D.} (1997).
  Sampling from the posterior distribution in generalized
linear mixed
models.
 \textit{Statist. Comput.} \textbf{7} 57--68.

\bibitem[\protect\citeauthoryear{Gamerman}{1998}]{Gamerman98}
\textsc{Gamerman, D.} (1998).
  Markov chain Monte Carlo for dynamic generalised linear models.
 \textit{Biometrika} \textbf{85} 215--227.
\MR{1627273}

\bibitem[\protect\citeauthoryear{Garofalo et~al.}{2009}]{Garofalo09}
\textsc{Garofalo, M., Nieus, T., Massobrio, P.} and \textsc{Martinoia, S.} (2009).
  Evaluation of the performance of information theory-based
methods and
cross-correlation to estimate the functional connectivity in cortical
networks.
 \textit{PLoS ONE} \textbf{4} e6482.

\bibitem[\protect\citeauthoryear{Godsill, Doucet and West}{2004}]{GDW04}
\textsc{Godsill, S., Doucet, A.} and \textsc{West, M.} (2004).
  Monte {C}arlo smoothing for non-linear time series.
 \textit{J. Amer. Statist. Assoc.} \textbf{99} 156--168.
\MR{2054295}

\bibitem[\protect\citeauthoryear{Gomez-Urquijo et~al.}{2000}]{Urquijo2000}
\textsc{Gomez-Urquijo, S.~M., Reblet, C., Bueno-Lopez, J.~L.} and \textsc{Gutierrez-Ibarluzea,
I.} (2000).
  Gabaergic neurons in the rabbit visual cortex: Percentage,
distribution and cortical projections.
 \textit{Brain Res.} \textbf{862} 171--179.

\bibitem[\protect\citeauthoryear{Greenberg, Houweling and Kerr}{2008}]{GreenbergKerr08}
\textsc{Greenberg, D.~S., Houweling, A.~R.} and \textsc{Kerr, J. N.~D.} (2008).
  Population imaging of ongoing neuronal activity in the visual cortex
of awake rats.
 \textit{Nat. Neurosci} \textbf{11} 749--751.

\bibitem[\protect\citeauthoryear{Gupta, Wang and Markram}{2000}]{Gupta00}
\textsc{Gupta, A., Wang, Y.} and \textsc{Markram, H.} (2000).
  Organizing principles for a diversity of gabaergic
interneurons and
synapses in the neocortex.
 \textit{Science} \textbf{287} 273--278.

\bibitem[\protect\citeauthoryear{Harris et~al.}{2003}]{HARR03}
\textsc{Harris, K., Csicsvari, J., Hirase, H., Dragoi, G.} and \textsc{Buzsaki, G.} (2003).
  Organization of cell assemblies in the hippocampus.
 \textit{Nature} \textbf{424} 552--556.

\bibitem[\protect\citeauthoryear{Hatsopoulos et~al.}{1998}]{HATS98}
\textsc{Hatsopoulos, N., Ojakangas, C., Paninski, L.} and \textsc{Donoghue, J.} (1998).
  Information about movement direction obtained by synchronous activity
of motor cortical neurons.
 \textit{PNAS} \textbf{95} 15706--15711.

\bibitem[\protect\citeauthoryear{Helmchen, Imoto and Sakmann}{1996}]{HelmchenSakmann96}
\textsc{Helmchen, F., Imoto, K.} and \textsc{Sakmann, B.} (1996).
 {Ca}$^{2+}$ buffering and action potential-evoked {Ca}$^{2+}$
signaling in dendrites of pyramidal neurons.
 \textit{Biophys. J.} \textbf{70} 1069--1081.

\bibitem[\protect\citeauthoryear{Ikegaya et~al.}{2004}]{IkegayaYuste04}
\textsc{Ikegaya, Y., Aaron, G., Cossart, R., Aronov, D., Lampl, I., Ferster,
D.} and \textsc{Yuste,~R.} (2004).
  Synfire chains and cortical songs: Temporal modules of cortical
activity.
 \textit{Science} \textbf{304} 559--564.

\bibitem[\protect\citeauthoryear{Ishwaran}{1999}]{Ishwaran99}
\textsc{Ishwaran, H.} (1999).
  Applications of hybrid {Monte Carlo to Bayesian} generalized linear
models: Quasicomplete separation and neural networks.
 \textit{J. Comput. Graph. Statist.} \textbf{8} 779--799.
\MR{1748967}

\bibitem[\protect\citeauthoryear{Iyer, Hoogland and Saggau}{2006}]{Iyer06}
\textsc{Iyer, V., Hoogland, T.~M.} and \textsc{Saggau, P.} (2006).
  Fast functional imaging of single neurons using random-access
multiphoton ({RAMP}) microscopy.
 \textit{J. Neurophysiol.} \textbf{95} 535--545.

\bibitem[\protect\citeauthoryear{Koch}{1999}]{Koch99}
\textsc{Koch, C.} (1999).
 \textit{Biophysics of Computation}.
  Oxford Univ. Press, Oxford.

\bibitem[\protect\citeauthoryear{Kulkarni and Paninski}{2007}]{KP06}
\textsc{Kulkarni, J.} and \textsc{Paninski, L.} (2007).
  Common-input models for multiple neural spike-train data.
 \textit{Network Comput.  Neural Syst.} \textbf{18} 375--407.

\bibitem[\protect\citeauthoryear{Lefort et~al.}{2009}]{Lefort2009}
\textsc{Lefort, S., Tomm, C., Floyd~Sarria, J.-C.} and \textsc{Petersen, C. C.~H.} (2009).
  The excitatory neuronal network of the c2 barrel column in mouse
primary somatosensory cortex.
 \textit{Neuron} \textbf{61} 301--316.

\bibitem[\protect\citeauthoryear{Lei et~al.}{2008}]{Shepard08}
\textsc{Lei, N., Watson, B., MacLean, J., Yuste, R.} and \textsc{Shepard, K.} (2008).
  A 256-by-256 cmos microelectrode array for extracellular stimulation
of acute brain slices.
In \textit{Proceedings to the International Solid-State Circuits
Conference}, ISSCC.

\bibitem[\protect\citeauthoryear{Li and Duan}{1989}]{LD89}
\textsc{Li, K.} and \textsc{Duan, N.} (1989).
  Regression analysis under link violation.
 \textit{Ann. Statist.} \textbf{17} 1009--1052.
\MR{1015136}

\bibitem[\protect\citeauthoryear{Litke et~al.}{2004}]{Litke2004}
\textsc{Litke, A., Bezayiff, N., Chichilnisky, E., Cunningham, W., Dabrowski, W.,
Grillo, A., Grivich, M., Grybos, P., Hottowy, P., Kachiguine, S.,
Kalmar, R.,
Mathieson, K., Petrusca, D., Rahman, M.} and \textsc{Sher, A.} (2004).
  What does the eye tell the brain? Development of a system for the
large scale recording of retinal output activity.
 \textit{IEEE Trans. Nucl. Sci.} \textbf{51} 1434--1440.

\bibitem[\protect\citeauthoryear{Livet et~al.}{2007}]{Brainbow07}
\textsc{Livet, J., Weissman, T., Kang, H., Draft, R., Lu, J., Bennis, R.,
Sanes, J.}
and \textsc{Lichtman, J.} (2007).
  Transgenic strategies for combinatorial expression of fluorescent
proteins in the nervous system.
 \textit{Nature} \textbf{450} 56--62.

\bibitem[\protect\citeauthoryear{Luczak et~al.}{2007}]{Harris07}
\textsc{Luczak, A., Bartho, P., Marguet, S., Buzsaki, G.} and \textsc{Harris, K.} (2007).
  Sequential structure of neocortical spontaneous activity in vivo.
 \textit{PNAS} \textbf{104} 347--352.

\bibitem[\protect\citeauthoryear{MacLean et~al.}{2005}]{MACLEAN05}
\textsc{MacLean, J., Watson, B., Aaron, G.} and \textsc{Yuste, R.} (2005).
  Internal dynamics determine the cortical response to thalamic
stimulation.
 \textit{Neuron} \textbf{48} 811--823.

\bibitem[\protect\citeauthoryear{McLachlan and Krishnan}{1996}]{McLachlanKrishnan96}
\textsc{McLachlan, G.} and \textsc{Krishnan, T.} (1996).
 \textit{The EM Algorithm and Extensions}.
  Wiley, New York.
\MR{2392878}

\bibitem[\protect\citeauthoryear{Meyer et~al.}{2002}]{Meyer02}
\textsc{Meyer, A.~H., Katona, I., Blatow, M., Rozov, A.} and \textsc{Monyer, H.} (2002).
 {In vivo labeling of parvalbumin-positive interneurons and analysis
of electrical coupling in identified neurons}.
 \textit{J.~Neurosci.} \textbf{22} 7055--7064.

\bibitem[\protect\citeauthoryear{Micheva and Smith}{2007}]{MichevaSmith07}
\textsc{Micheva, K.} and \textsc{Smith, S.} (2007).
  Array tomography: A new tool for imaging the molecular architecture
and ultrastructure of neural circuits.
 \textit{Neuron} \textbf{55} 25--36.

\bibitem[\protect\citeauthoryear{Mishchenko}{2009}]{Mishchenko2009}
\textsc{Mishchenko, Y.} (2009).
  Strategies for identifying exact structure of neural circuits with
broad light microscopy connectivity probes.
Preprint. Available at \url{http://precedings.nature.com/documents/2669/version/2}.

\bibitem[\protect\citeauthoryear{Mishchenko et~al.}{2009}]{Mishchenko2009b}
\textsc{Mishchenko, Y., Spacek, J., Mendenhall, J., Chklovskii, D.} and \textsc{Harris, K.~M.}
(2009).
  Reconstruction of hippocampal {CA1} neuropil at nanometer resolution
reveals disordered packing of processes and dependence of synaptic
connectivity on local environment and dendritic caliber. To appear.

\bibitem[\protect\citeauthoryear{Neal, Beal and Roweis}{2003}]{NBR03}
\textsc{Neal, R., Beal, M.} and \textsc{Roweis, S.} (2003).
  Inferring state sequences for non-linear systems with
embedded hidden
{M}arkov models.
In \textit{NIPS} \textbf{16} 401--408. MIT Press, Cambridge.

\bibitem[\protect\citeauthoryear{Ng}{2004}]{NG04}
\textsc{Ng, A.} (2004).
  Feature selection, {L}$_1$ vs. {L}$_2$ regularization, and rotational
invariance. In \textit{Proceedings of the Twenty-First International Conference on
Machine Learning}. \textit{ICML} 21.

\bibitem[\protect\citeauthoryear{Nguyen et~al.}{2001}]{NguyenParker01}
\textsc{Nguyen, Q.~T., Callamaras, N., Hsieh, C.} and \textsc{Parker, I.} (2001).
  Construction of a~two-photon microscope for video-rate {Ca}$^{2+}$
imaging.
 \textit{Cell Calcium} \textbf{30} 383--393.

\bibitem[\protect\citeauthoryear{Nikolenko, Poskanzer and Yuste}{2007}]{Vovan07}
\textsc{Nikolenko, V., Poskanzer, K.} and \textsc{Yuste, R.} (2007).
  Two-photon photostimulation and imaging of neural circuits.
 \textit{Nature Methods} \textbf{4} 943--950.

\bibitem[\protect\citeauthoryear{Nikolenko et~al.}{2011}]{Nikolenko08}
\textsc{Nikolenko, V., Watson, B., Araya, R., Woodruff, A., Peterka, D.} and
\textsc{Yuste, R.}
(2011).
 {SLM} microscopy: Scanless two-photon imaging and photostimulation
using spatial light modulators.
 \textit{Frontiers in Neural Circuits}. To appear.
\href{http://dx.doi.org/10.3389/neuro.04.005.2008}{DOI:10.3389/neuro.04.005.2008}.\vadjust{\goodbreak}

\bibitem[\protect\citeauthoryear{Nykamp}{2005}]{Nykamp05}
\textsc{Nykamp, D.~Q.} (2005).
  Revealing pairwise coupling in linear--nonlinear networks.
 \textit{SIAM J. Appl. Math.} \textbf{65} 2005--2032.
\MR{2177736}

\bibitem[\protect\citeauthoryear{Nykamp}{2007}]{NYK06}
\textsc{Nykamp, D.~Q.} (2007).
  A mathematical framework for inferring connectivity in probabilistic
neuronal networks.
 \textit{Math. Biosci.} \textbf{205} 204--251.
\MR{2295044}



\bibitem[\protect\citeauthoryear{Ohki et~al.}{2005}]{OHKI05}
\textsc{Ohki, K., Chung, S., Ch'ng, Y., Kara, P.} and \textsc{Reid, C.} (2005).
  Functional imaging with cellular resolution reveals precise
micro-architecture in visual cortex.
 \textit{Nature} \textbf{433} 597--603.

\bibitem[\protect\citeauthoryear{Paninski}{2004}]{PAN04c}
\textsc{Paninski, L.} (2004).
  Maximum likelihood estimation of cascade point-process neural
encoding models.
 \textit{Network Comput. Neural Syst.} \textbf{15} 243--262.

\bibitem[\protect\citeauthoryear{Paninski et~al.}{2009}]{Pan08b}
\textsc{Paninski, L., Ahmadian, Y., Ferreira, D., Koyama, S., Rahnama, K.,
Vidne, M.,
Vogelstein, J.} and \textsc{Wu, W.} (2009).
  A new look at state-space models for neural data.
 \textit{J. Comput. Neurosci.} To appear.

\bibitem[\protect\citeauthoryear{Paninski et~al.}{2004}]{PAN03d}
\textsc{Paninski, L., Fellows, M., Shoham, S., Hatsopoulos, N.} and \textsc{Donoghue, J.}
(2004).
  Superlinear population encoding of dynamic hand trajectory in primary
motor cortex.
 \textit{J. Neurosci.} \textbf{24} 8551--8561.

\bibitem[\protect\citeauthoryear{Petersen and Sakmann}{2000}]{PetersenSakmann00}
\textsc{Petersen, C.~C.} and \textsc{Sakmann, B.} (2000).
  The excitatory neuronal network of rat layer 4 barrel cortex.
 \textit{J. Neurosci.} \textbf{20} 7579--7586.

\bibitem[\protect\citeauthoryear{Petrusca et~al.}{2007}]{Petrusca07}
\textsc{Petrusca, D., Grivich, M.~I., Sher, A., Field, G.~D., Gauthier, J.~L.,
Greschner, M., Shlens, J., Chichilnisky, E.~J.} and \textsc{Litke, A.~M.} (2007).
  Identification and characterization of a {Y}-like primate retinal
ganglion cell type.
 \textit{J. Neurosci.} \textbf{27} 11019--11027.

\bibitem[\protect\citeauthoryear{Pillow et~al.}{2008}]{PILL07}
\textsc{Pillow, J., Shlens, J., Paninski, L., Sher, A., Litke, A.,
Chichilnisky, E.}
and \textsc{Simoncelli, E.} (2008).
  Spatiotemporal correlations and visual signaling in a complete
neuronal population.
 \textit{Nature} \textbf{454} 995--999.

\bibitem[\protect\citeauthoryear{Plesser and Gerstner}{2000}]{PG00}
\textsc{Plesser, H.} and \textsc{Gerstner, W.} (2000).
  Noise in integrate-and-fire neurons: From stochastic input to escape
rates.
 \textit{Neural Comput.} \textbf{12} 367--384.

\bibitem[\protect\citeauthoryear{Rabiner}{1989}]{RAB89}
\textsc{Rabiner, L.} (1989).
  A tutorial on hidden {M}arkov models and selected
applications in
speech recognition.
 \textit{Proc. IEEE} \textbf{77} 257--286.

\bibitem[\protect\citeauthoryear{Ramon~y Cajal}{1904}]{RamonyCajal04}
\textsc{Ramon~y Cajal, S.} (1904).
 \textit{La Textura del Sistema Nerviosa del Hombre y los Vertebrados}.
  Moya, Madrid.

\bibitem[\protect\citeauthoryear{Ramon~y Cajal}{1923}]{RamonyCajal23}
\textsc{Ramon~y Cajal, S.} (1923).
 \textit{Recuerdos de mi vida: Historia de mi labor cientifica}.
  Alianza Editorial, Madrid.

\bibitem[\protect\citeauthoryear{Reddy et~al.}{2008}]{ReddySaggau08}
\textsc{Reddy, G., Kelleher, K., Fink, R.} and \textsc{Saggau, P.} (2008).
 {Three-dimensional random access multiphoton microscopy for
functional imaging of neuronal activity}.
 \textit{Nat. Neurosci.} \textbf{11} 713--720.

\bibitem[\protect\citeauthoryear{Reyes et~al.}{1998}]{Reyes98}
\textsc{Reyes, A., Lujan, R., Rozov, A., Burnashev, N., Somogyi, P.} and
\textsc{Sakmann, B.}
(1998).
  Target-cell-specific facilitation and depression in neocortical
circuits.
 \textit{Nat. Neurosci.}  \textbf{1} 279--285.

\bibitem[\protect\citeauthoryear{Rigat, de~Gunst and van Pelt}{2006}]{Rigat06}
\textsc{Rigat, F., de~Gunst, M.} and \textsc{van Pelt, J.} (2006).
  Bayesian modelling and analysis of spatio-temporal neuronal networks.
 \textit{Bayesian Anal.} \textbf{1} 733--764.
\MR{2282205}

\bibitem[\protect\citeauthoryear{Robert and Casella}{2005}]{RC05}
\textsc{Robert, C.} and \textsc{Casella, G.} (2005).
 \textit{Monte {C}arlo Statistical Methods}.
  Springer, New York.

\bibitem[\protect\citeauthoryear{Roxin, Hakim and Brunel}{2008}]{Roxin08}
\textsc{Roxin, A., Hakim, V.} and \textsc{Brunel, N.} (2008).
  The statistics of repeating patterns of cortical activity can be
reproduced by a model network of stochastic binary neurons.
 \textit{J. Neurosci.} \textbf{28} 10734--10745.

\bibitem[\protect\citeauthoryear{Salome et~al.}{2006}]{SalomeBourdieu06}
\textsc{Salome, R., Kremer, Y., Dieudonne, S., Leger, J.-F., Krichevsky, O.,
Wyart, C.,
Chatenay, D.} and \textsc{Bourdieu, L.} (2006).
  Ultrafast random-access scanning in two-photon microscopy using
acousto-optic deflectors.
 \textit{J. Neurosci. Methods} \textbf{154} 161--174.

\bibitem[\protect\citeauthoryear{Santhanam et~al.}{2006}]{Santhanam06}
\textsc{Santhanam, G., Ryu, S.~I., Yu, B.~M., Afshar, A.} and \textsc{Shenoy, K.~V.} (2006).
  A~high-performance brain-computer interface.
 \textit{Nature} \textbf{442} 195--198.

\bibitem[\protect\citeauthoryear{Sayer, Friedlande and Redman}{1990}]{Sayer1990}
\textsc{Sayer, R.~J., Friedlander, M.~J.} and \textsc{Redman, S.~J.} (1990).
  The time course and amplitude of epsps evoked at synapses between
pairs of {CA3/CA1} neurons in the hippocampal slice.
 \textit{J.~Neurosci.} \textbf{10} 826--836.

\bibitem[\protect\citeauthoryear{Segev et~al.}{2004}]{Berry2004}
\textsc{Segev, R., Goodhouse, J., Puchalla, J.} and \textsc{Berry, M.} (2004).
  Recording spikes from a large fraction of the ganglion cells
in a
retinal patch.
 \textit{Nat. Neurosci.} \textbf{7} 1154--1161.

\bibitem[\protect\citeauthoryear{Shumway and Stoffer}{2006}]{ShumwayStoffer06}
\textsc{Shumway, R.} and \textsc{Stoffer, D.} (2006).
 \textit{Time Series Analysis and Its Applications}.
  Springer, New York.
\MR{2228626}

\bibitem[\protect\citeauthoryear{Song et~al.}{2005}]{Song2005}
\textsc{Song, S., Sjostrom, P.~J., Reiql, M., Nelson, S.} and \textsc{Chklovskii, D.~B.} (2005).
  Highly nonrandom features of synaptic connectivity in local cortical
circuits.
 \textit{PLoS Biol.} \textbf{3} e68.

\bibitem[\protect\citeauthoryear{Stein et~al.}{2004}]{Stein04}
\textsc{Stein, R.~B., Weber, D.~J., Aoyagi, Y., Prochazka, A., Wagenaar, J. B.~M.,
Shoham, S.} and \textsc{Normann, R.~A.} (2004).
 {Coding of position by simultaneously recorded sensory
neurones in
the cat dorsal root ganglion}.
 \textit{J. Physiol.} \textbf{560} 883--896.

\bibitem[\protect\citeauthoryear{Stevenson et~al.}{2008}]{Stevenson08}
\textsc{Stevenson, I., Rebesco, J., Hatsopoulos, N., Haga, Z., Miller, L.} and
\textsc{Koerding, K.} (2008).
  Inferring network structure from spikes.
In \textit{Statistical Analysis of Neural Data Meeting}.

\bibitem[\protect\citeauthoryear{Stevenson et~al.}{2009}]{Stevenson2009}
\textsc{Stevenson, I.~H., Rebesco, J.~M., Hatsopoulos, N.~G., Haga, Z., Miller, L.~E.}
and \textsc{Kording, K.~P.} (2009).
  Bayesian inference of functional connectivity and network structure
from spikes.
 \textit{IEEE Trans. Neural Syst. Rehab.} \textbf{17} 203--213.

\bibitem[\protect\citeauthoryear{Stosiek et~al.}{2003}]{StosiekKonnerth03}
\textsc{Stosiek, C., Garaschuk, O., Holthoff, K.} and \textsc{Konnerth, A.} (2003).
  In vivo two-photon calcium imaging of neuronal networks.
 \textit{Proc. Natl. Acad. Sci. USA} \textbf{100} 7319--7324.

\bibitem[\protect\citeauthoryear{Szobota et~al.}{2007}]{SzobotaIsacoff07}
\textsc{Szobota, S., Gorostiza, P., Del~Bene, F., Wyart, C., Fortin, D.~L., Kolstad,
K.~D., Tulyathan, O., Volgraf, M., Numano, R., Aaron, H.~L., Scott, E.~K.,
Kramer, R.~H., Flannery, J., Baier, H., Trauner, D.} and \textsc{Isacoff, E.~Y.}
(2007).
  Remote control of neuronal activity with a light-gated glutamate
receptor.
 \textit{Neuron} \textbf{54} 535--545.

\bibitem[\protect\citeauthoryear{Thompson, Girdlestone and West}{1988}]{Thompson88}
\textsc{Thompson, A., Girdlestone, D.} and \textsc{West, D.} (1988).
  Voltage-dependent currents prolong single-axon postsynaptic
potentials in layer {III} pyramidal neurons in rat neocortical slices.
 \textit{J.~Neurophysiol.} \textbf{60} 1896--1907.

\bibitem[\protect\citeauthoryear{Tibshirani}{1996}]{Tibs96}
\textsc{Tibshirani, R.} (1996).
  Regression shrinkage and selection via the Lasso.
 \textit{J.~Roy. Statist. Soc. Ser. B}
\textbf{58} 267--288.
\MR{1379242}

\bibitem[\protect\citeauthoryear{Tipping}{2001}]{TIP01}
\textsc{Tipping, M.} (2001).
  Sparse {B}ayesian learning and the relevance vector machine.
 \textit{J. Mach. Learn. Res.} \textbf{1} 211--244.
\MR{1875838}

\bibitem[\protect\citeauthoryear{Truccolo et~al.}{2005}]{TRUC05}
\textsc{Truccolo, W., Eden, U., Fellows, M., Donoghue, J.} and \textsc{Brown, E.} (2005).
  A~point process framework for relating neural spiking
activity to
spiking history, neural ensemble and extrinsic covariate effects.
 \textit{J. Neurophysiol.} \textbf{93} 1074--1089.

\bibitem[\protect\citeauthoryear{Tsien}{1989}]{Tsien89}
\textsc{Tsien, R.~Y.} (1989).
  Fluorescent probes of cell signaling.
 \textit{Ann. Rev. Neurosci.} \textbf{12} 227--253.

\bibitem[\protect\citeauthoryear{Vakorin, Krakovska and Mcintosh}{2009}]{Vakorin09}
\textsc{Vakorin, V.~A., Krakovska, O.~A.} and \textsc{Mcintosh, A.~R.} (2009).
  Confounding effects of indirect connections on causality estimation.
 \textit{J. Neurosci. Methods} \textbf{184} 152--160.

\bibitem[\protect\citeauthoryear{Vidne et~al.}{2009}]{Vidne08}
\textsc{Vidne, M., Kulkarni, J., Ahmadian, Y., Pillow, J., Shlens, J., Chichilnisky,
E., Simoncelli, E.} and \textsc{Paninski, L.} (2009).
  Inferring functional connectivity in an ensemble of retinal ganglion
cells sharing a common input.
 In \textit{Computational and Systems Neuroscience
(COSYNE09)}.

\bibitem[\protect\citeauthoryear{Vogels and Abbott}{2005}]{Vogels05}
\textsc{Vogels, T.} and \textsc{Abbott, L.~F.} (2005).
 {Signal propagation and logic gating in networks of
integrate-and-fire neurons}.
 \textit{J. Neurosci.} \textbf{25} 10786--10795.

\bibitem[\protect\citeauthoryear{Vogelstein et~al.}{2008}]{Vogelstein08}
\textsc{Vogelstein, J., Babadi, B., Watson, B., Yuste, R.} and \textsc{Paninski, L.} (2008).
  Fast nonnegative deconvolution via tridiagonal interior-point
methods, applied to calcium fluorescence data.
In \textit{Statistical Analysis of Neural Data (SAND) Conference}.

\bibitem[\protect\citeauthoryear{Vogelstein et~al.}{2009}]{Vogelstein2009}
\textsc{Vogelstein, J., Watson, B., Packer, A., Jedynak, B., Yuste, R.} and \textsc{Paninski,
L.} (2009).
  Spike inference from calcium imaging using sequential monte carlo
methods.
 \textit{Biophys. J.} \textbf{97} 636--655.

\bibitem[\protect\citeauthoryear{Wallace et~al.}{2008}]{WallaceHasan08}
\textsc{Wallace, D., zum Alten~Borgloh, S., Astori, S., Yang, Y., Bausen, M., Kugler,
S., Palmer, A., Tsien, R., Sprengel, R., Kerr, J., Denk, W.} and
\textsc{Hasan, M.}
(2008).
 {Single-spike detection in vitro and in vivo with a genetic Ca$^{2+}$
sensor}.
 \textit{Nat. Methods} \textbf{5} 797--804.

\bibitem[\protect\citeauthoryear{Yaksi and Friedrich}{2006}]{YaksiFriedrich06}
\textsc{Yaksi, E.} and \textsc{Friedrich, R.~W.} (2006).
  Reconstruction of firing rate changes across neuronal
populations by
temporally deconvolved {Ca$^{2+}$} imaging.
 \textit{Nat. Methods} \textbf{3} 377--383.

\bibitem[\protect\citeauthoryear{Yasuda et~al.}{2004}]{Yasuda2004}
\textsc{Yasuda, R., Nimchinsky, E.~A., Scheuss, V., Pologruto, T.~A., Oertner, T.~G.,
Sabatini, B.~L.} and \textsc{Svoboda, K.} (2004).
  Imaging calcium concentration dynamics in small neuronal
compartments.
 \textit{Sci. STKE} \textbf{219} 15.

\bibitem[\protect\citeauthoryear{Yuste et~al.}{2006}]{ImagingManual}
\textsc{Yuste, R., Konnerth, A., Masters, B. et~al.} (2006).
\textit{{Imaging in Neuroscience and Development, A Laboratory
Manual}}. Oxford, New York.

\end{thebibliography}
\end{document}